\documentclass[reprint, amsmath, amssymb, aps]{revtex4-2}

\usepackage{graphicx}
\usepackage{dcolumn}
\usepackage{bm}
\usepackage{subfigure}
\usepackage[colorlinks,linkcolor=blue,citecolor=blue]{hyperref}
\usepackage{comment}
\usepackage{pgfplots}
\pgfplotsset{compat = newest}

\graphicspath{ {images/} }

\begin{document}

\preprint{APS/123-QED}

\title{Mapping Sandpiles to Complex Networks}

\author{A. Shoja-Daliklidash}
\affiliation{Department of Physics, University of Mohaghegh Ardabili, P.O. Box 179, Ardabil, Iran}
\author{M. Nattagh-Najafi}
\email{morteza.nattagh@gmail.com}
\affiliation{Department of Physics, University of Mohaghegh Ardabili, P.O. Box 179, Ardabil, Iran}
\author{N. Sepehri-Javan}
\affiliation{Department of Physics, University of Mohaghegh Ardabili, P.O. Box 179, Ardabil, Iran}


\begin{abstract}
In this paper, we address a longstanding challenge in self-organized criticality (SOC) systems: establishing a connection between sandpiles and complex networks. Our approach employs a similarity-based transfer function characterized by two parameters, $\mathcal{R}=(r_1, r_2)$. Here, $r_1$ quantifies the similarity of local activities, while $r_2$ governs the filtration process used to convert a weighted network into a binary one. We reveal that the degree centrality distribution of the resulting network follows a generalized Gamma distribution (GGD), which transitions to a power-law distribution under specific conditions. The GGD exponents, estimated numerically, exhibit a dependency on $\mathcal{R}$. Notably, while both decreasing $r_1$ and $r_2$ lead to denser networks, $r_2$ plays a more significant role in influencing network density. Furthermore, the Shannon entropy is observed to decrease linearly with increasing $r_2$, whereas its variation with $r_1$ is more gradual. An analytical expression for the Shannon entropy is proposed. To characterize the network structure, we investigate the clustering coefficient ($cc$), eigenvalue centrality ($e$), closeness centrality ($c$), and betweenness centrality ($b$). The distributions of $cc$, $e$, and $c$ exhibit peaked profiles, while $b$ displays a power-law distribution over a finite interval of $k$. Additionally, we explore correlations between the exponents and identify a specific parameter regime of $\mathcal{R}$ and $k$ where the $e-k$, $c-k$, and $b-k$ correlations become negative.
\end{abstract}

\maketitle

\section{Introduction}
The sandpile model, introduced by Bak, Tang, and Weisenfeld (BTW) in 1987, serves as a foundational example of self-organized criticality (SOC), a phenomenon observed in various natural systems. SOC systems naturally evolve toward a critical state without requiring external tuning or control \cite{bak1987self}. In the BTW model, energy is gradually added to the system, triggering localized relaxations in both time and space. These relaxations lead to a cascade of activities, termed \textit{avalanches}, whose sizes exhibit a distribution of power law across all spatial dimensions~\cite{kadanoff1989scaling}. Similar power-law behavior and scaling relationships are also observed in other statistical measures, including avalanche mass, duration, gyration radius, and external loop length ~\cite{Jensen_1989, dhar1990self,Najafi2012Moghimi,manna1999sandpile}. The BTW model was initially developed to account for $1/f$ noise~\cite{bak1987self}. However, it was soon recognized by the model's proponents that the phenomenon extended to other power-law behaviors of $f$\cite{bak1988self, tang1988critical}. Dhar later discovered the Abelian structure of sandpiles, leading to the development of the Abelian Sandpile Model (ASM)\cite{dhar1990self, redig2012abelian}. This mathematical framework revealed connections to the Potts model\cite{welsh2000potts}, directed percolation~\cite{pastor2000universality}, restricted solid-on-solid (RSOS) models\cite{De_Luca_2013}, the voter model~\cite{ali1995structure}, and spanning trees~\cite{levine2011sandpile}, significantly advancing the understanding of complex systems~\cite{dhar1999abelian}. The ASM is now a powerful tool for exploring critical behaviors and temporal dynamics in such systems~\cite{dhar2006theoretical}. Various extensions and adaptations of the sandpile model have also been studied, including fixed-energy sandpiles~\cite{manna1999sandpile}, non-stationary sandpiles~\cite{manna2023nonstationary}, diffusive sandpiles~\cite{Najafi_2020}, non-conservative models for earthquakes~\cite{Lise2001earthquake}, invasion sandpiles~\cite{Najafi_2020}, sandpiles modeling fluid propagation~\cite{NAJAFI2016102}, and two-dimensional cuts of $2d$ sandpiles~\cite{dashti2015statistical}.

Sandpile models have been studied on various supports, including scale-free networks, random networks, correlated~\cite{saberi2015recent} and uncorrelated~\cite{de2018percolation} percolation lattices, and random lattices with long-range interactions~\cite{li2016many}. On supports with a high number of local interactions, such as scale-free networks, mean-field theory is expected to provide a better approximation~\cite{barabasi1999mean}. For example, Kim developed a mean-field solution for the BTW model, revealing that the avalanche size distribution follows a power law~\cite{christensen1993sandpile}. These studies gain additional significance when interpreted as dynamic models of neuronal activity. In particular, the integrate-and-fire model exhibits similarities to the sandpile dynamics, such as the presence of a threshold value beyond which neurons fire~\cite{moosavi2014mean, rahimi2021role}.\\

The integration of sandpile models with complex networks marks a pivotal development in self-organized criticality (SOC) research, as it facilitates uncovering hidden dynamics within SOC systems by leveraging the well-established topological and statistical properties of complex networks. This approach finds diverse applications in areas such as biological ecosystems, social interactions, technological infrastructures, and information networks~\cite{strogatz2001exploring,landi2018complexity,frasco2014spatially,sole2004information}. Complex networks provide a robust framework for modeling and analyzing the organization of complex systems, elucidating principles that govern collective behavior, resilience, and adaptability. Their geometric structure facilitates the study of system dynamics, with networks featuring small-world or scale-free topologies exhibiting characteristics such as high clustering and resilience to random failures. These properties profoundly affect processes like information spread and influence propagation within the network~\cite{newman2018networks}. Early studies investigated the behavior of sandpile models on scale-free, small-world, and geographical networks~\cite{pan2007sandpile, bhaumik2018stochastic, lelarge2012diffusion}, revealing that properties like node degree, which serves as a threshold for avalanches, significantly impact avalanche dynamics and network stability~\cite{lee2004branching, lee2015forest, goh2003sandpile}. The BTW model has also been studied on random networks with limited-range interactions~\cite{najafi2014bak} and various other structures~\cite{najafi2018statistical, najafi2018sandpile, najafi2020avalanches, najafi2020geometry, cheraghalizadeh2017mapping, najafi2016bak}. Notably, the impact of varying the degree exponent $\textit{gamma}$ of scale-free networks on the temporal and spatial scaling exponents of avalanches has been explored~\cite{goh2003sandpile}, underscoring the critical role of network topology in influencing SOC behaviors.

The mapping of sandpile models onto complex networks has extended to various practical domains, offering valuable insights. In seismology, the recurrence time distribution of earthquakes and aftershocks has been shown to exhibit self-organized criticality, suggesting that SOC principles may aid in understanding and potentially predicting earthquake behavior~\cite{corral2004long, najafi2020avalanches2}. Similarly, in neuronal networks, background activity has been found to enable large networks of neurons to sustain a critical state, with network size and activity levels playing a crucial role in triggering seizure-like behaviors~\cite{hesse2014self}. Recent studies have further demonstrated the emergence of novel critical phenomena in sandpile models mapped onto complex networks, highlighting the impact of network structures on the scaling properties and stability of these systems, thereby uncovering new aspects of critical dynamics~\cite{fazli2022emergence}. Liu and colleagues investigated cascading phenomena in networks, demonstrating how structural properties govern the spread of disturbances across systems~\cite{cajueiro2010controlling, liu2012cascading}. In a related vein, studies on the Manna Model in complex networks, particularly Barabási-Albert networks, revealed a phase transition to an absorbing state. Interestingly, the critical exponents observed in these networks resemble those of directed percolation rather than those found in regular networks, highlighting the distinct dynamics introduced by complex network structures~\cite{bhaumik2018conserved}.\\

Despite significant advancements, challenges persist in linking sandpile models to complex networks. A major issue lies in the methods used to connect nodes, which can lead to debates over the validity of these mappings. For instance, falsely linking independent events based on temporal succession or the complications introduced by thresholding can significantly alter network topology and geometry. Furthermore, many mappings are designed for specific problems, complicating cross-system comparisons. As a result, while complex networks offer a promising framework for studying SOC systems, critical questions remain, particularly regarding the influence of thresholding and network geometry on the critical behavior of these models. In this paper, we propose a systematic method for mapping sandpile models onto complex networks, with a specific focus on thresholding and its effects on network properties. We show that this mapping adheres to specific laws, offering a general framework applicable to a wide range of SOC systems. Our findings provide valuable insights into how network structures shape SOC behaviors, emphasizing the importance of a structured approach to effectively map SOC systems onto complex networks.\\

The paper is organized as follows: In the next section, we introduce the BTW model and present the motivations behind this study. Section~\ref{SEC:Mapping} explains how we map this SOC system to a complex network. To make a binary network, we need a filtration process, which is explained in SEC.~\ref{SEC:filtration}. Section~\ref{SEC:Results} is devoted to explaining the structure of network, which contains two main parts: the local measurements (SEC.~\ref{SEC:local}), and the global measures (SEC.~\ref{SEC:global}). The local measures that we study are degree centrality (SEC.~\ref{SEC:degree}), Shannon entropy (SEC.~\ref{SEC:shannon}), clustering coefficient (SEC.~\ref{SEC:cc}), and eigenvector centrality (SEC.~\ref{SEC:Eigen}). The global measures include the closeness centrality (SEC.~\ref{SEC:closeness}) and the betweenness centrality (SEC.~\ref{SEC:betweenness}). We conclude the paper in SEC.~\ref{SEC:concl}.
\section{THE MODEL AND MOTIVATION}\label{SEC:model}
There is a growing interest in investigating self-organized critical (SOC) systems within the framework of complex networks (CNs). Although prior studies have explored these systems, many have largely overlooked the internal dynamics, particularly in sandpile models~\cite{levine2011sandpile}. This research addresses this gap by focusing on the micro-level interactions that occur during avalanche events, aiming to uncover previously unrecognized insights into SOC behavior.

\subsection{BAK-TANG-WEISENFELD MODEL}~\label{SEC:BTW}
The Bak–Tang–Wiesenfeld (BTW) model is a cellular automaton that captures the essence of self-organized criticality (SOC) in a sandpile system. In this model, each site on a lattice can hold a specific number of sand grains, and when the number exceeds a threshold, the site topples, redistributing its grains to neighboring sites. This redistribution can trigger a cascading series of topplings, referred to as an avalanche. Despite its simplicity, the BTW model is a powerful tool for studying diverse phenomena, including earthquakes, forest fires, and financial markets~\cite{newman2018networks}. As a prime example of a self-organized system, it demonstrates complex behavior arising spontaneously without external control.\\

We define the model on an $L \times L$ square lattice, where sand grains are randomly distributed, forming a height field $h$. Each site can hold a maximum height of $2d$, where $d$ represents the dimensionality of the system. The system is open, allowing for the addition or dissipation of energy. The dynamics of the model proceed as follows:

\begin{enumerate}
    \item A random site $i$ is selected, and a grain is added, increasing its height by 1: $h(i) \rightarrow h(i) + 1$.
    
    \item If the new height is less than the critical value $h_c = 2d$, another site is chosen for adding a grain.
    
    \item If the height exceeds the critical value, the site becomes unstable and topples. During the toppling, the height of the site decreases by $2d$, and each neighboring site's grain count increases by 1, conserving the total number of grains.
\end{enumerate}
The toppling process is described by:
\begin{equation}
 h(i) \rightarrow h(i) - \Delta_{i,j} 
\end{equation}
where
\begin{equation}
    \Delta_{i,j} = \begin{cases}
-1 & \text{if } i \text{ and } j \text{ are neighbors} \\
2d & \text{if } i = j \\
0 & \text{otherwise}
\end{cases} 
\end{equation}
The matrix $\Delta$ is known as the discrete Laplacian matrix. When a site topples, its neighbors may also become unstable and topple, triggering a chain reaction. This process continues until the system stabilizes, with all sites remaining below the critical height. The sequence of toppled sites during this process is referred to as an avalanche. The system can exhibit two types of configurations: transient and recurrent. Transient configurations occur early in the evolution and do not repeat, while recurrent configurations appear in the steady state, where the energy input and output are balanced. The total number of recurrent states is given by $\det(\Delta)$. For further details, refer to \cite{dhar2006theoretical}. A key property of this model is its self-organization into a critical state.

We implemented a detailed simulation of the Bak-Tang-Wiesenfeld (BTW) sandpile dynamics on a $2d$ lattice with $L = 100$, resulting in a total of 10,000 sites. The number of toppling events for each site was tracked, and this information was recorded in a toppling matrix to construct the complex networks (CNs).

\subsection{MAPPING TO COMPLEX NETWORKS}~\label{SEC:Mapping}
Over the four decades since the introduction of the sandpile concept, many aspects of these systems remain poorly understood, with the precise identification of universality classes being one of the key challenges. As a quintessential example of self-organized criticality (SOC), sandpile systems exhibit complex behavior during avalanche events; however, the local interactions between nodes have yet to be fully explored. This study addresses this gap by developing an enhanced model that incorporates these micro-level dynamics, providing new insights into the internal structure of sandpile systems. This approach enables the investigation of how local interactions impact global behavior, thereby offering a more profound understanding of the mechanisms that govern SOC systems. Our findings not only advance the comprehension of SOC but also have wider implications for predicting and controlling CNs dynamics. These results contribute to the development of a more comprehensive theory of self-organized criticality in networked systems. To facilitate this, we use the following mapping to relate sandpiles to CNs:
\begin{equation}
\text{W} = \log\left(\frac{L}{d_{xy}}\right) \cdot \exp\left(-r_{1} \cdot \frac{|n_{x} - n_{y}|}{n_{\text{max}}}\right)
\label{eq:Mapping_Network}
\end{equation}
where \( d_{xy} \) is the Euclidean distance between the sites, \( n_{\text{max}} \) is the maximum number of topplings recorded in the toppling matrix, \( r_{1} \) is control parameters, with ranging from 0.1 to 1 in steps of 0.1. 

\subsection{FILTRATION PROCESS}~\label{SEC:filtration}
In weighted connection networks, there exist weak links, or nodes that are weakly connected to others. The removal of such links or nodes shapes the connection network into its most relevant form for geometric or dynamical analysis. This process signifies diminishing the connections between sites with low toppling rates and those that are very distant from each other in the BTW model. This removal procedure is referred to as the filtration (sparsification) process, which defines a threshold beyond which the connection or interaction is removed. Such a filtration process naturally occurs in real data analysis when considering lower cutoffs in interactions. An important question arises: How do the results depend on such a threshold, and how does the system respond to the filtration process? In some cases, there is hope to find a pattern of behavior as a function of this threshold, enabling one to extrapolate the results.\\

When applying a threshold to convert a dense network into a sparse one, several changes occur across various network features, including the distribution of degree, eigenvalues, clustering coefficients, betweenness centrality, and closeness centrality. In the dense regime, the degree of nodes is high. As the threshold increases, weaker connections begin to be removed, prompting a transformation of the degree distribution. In non-scale-free networks with a peaked degree distribution, the peak position ($\bar{k}(a_{\text{th}})$, where $a_{\text{th}}$ is the considered threshold) is expected to be high in the dense regime, while diluting the network (increasing $a_{\text{th}}$) is accompanied by decreasing $\bar{k}$. In some circumstances, this transition may shift towards a bimodal pattern, where a large number of nodes with very low degrees have lost their connections, alongside a smaller set of highly interconnected nodes forming the core of the network. These core nodes become pivotal, acting as hubs that significantly influence the overall connectivity and dynamics of the network.

The effects of sparsification extend to the behaviors of eigenvalues and eigenvectors, which are crucial for understanding the network's structure. As sparsification occurs, low eigenvalues tend to increase due to a higher proportion of nodes becoming weakly connected or isolated. Conversely, high eigenvalues emerge from clusters of tightly connected, influential nodes that maintain their strong interconnections, leading to a concentration of centrality within these clusters. Moreover, intermediate eigenvalues decrease as many nodes with moderate connectivity lose their connections, either becoming more isolated or merging into the dominant clusters, emphasizing the increasingly polarized structure of the network. Additionally, the clustering coefficient diminishes as the network loses many of its local clusters, signaling a transition from a tightly-knit structure to one marked by greater dispersion. Changes in betweenness centrality are also significant; as the network becomes sparse, the betweenness centrality of certain key nodes increases, making these nodes critical connectors between different components of the network. This facilitates communication and influence across disparate sections. Overall, the transition from a dense to a sparse network results in a hierarchical structure dominated by a few influential hubs. This structural evolution profoundly alters the network's dynamics, reflecting the interplay between sparsity and the removal of weaker edges.

In this paper the following threshold is employed to filter out insignificant edges:
\begin{equation}
W_{th}(r_2) = r_2 \cdot w_{\text{max}} + (1 - r_2) \cdot w_{\text{avg}}
\end{equation}
where \( r_2 \) is a parameter ranging from 0 to 0.9 in steps of 0.1, \( w_{\text{max}} \) is the maximum edge weight, and \( w_{\text{avg}} \) is the average edge weight. In the following sections, we parameterize the networks using the couple $\mathcal{R}\equiv (r_1, r_2)$, and represent the network of size $N$ by $\mathcal{G}_{\mathcal{R}}(N)$. Note that $N$ corresponds to the avalanche mass of the original BTW model. It is important to note that $\mathcal{R}$ plays the role of a sparsification parameter in this paper. While both $r_1$ and $r_2$ contribute to this sparsification, they do so in different ways: $r_2$ addresses standard thresholding, while $r_1$ influences the \textit{similarity} of active sites. Specifically, as $r_1$ increases, the weight of the link between two nodes with different numbers of topplings decreases, thereby reducing the probability of their connection.

\section{THE NETWORK STRUCTURE}~\label{SEC:Results}
\begin{figure*}[t]
    \centering
    \includegraphics[width=\linewidth]{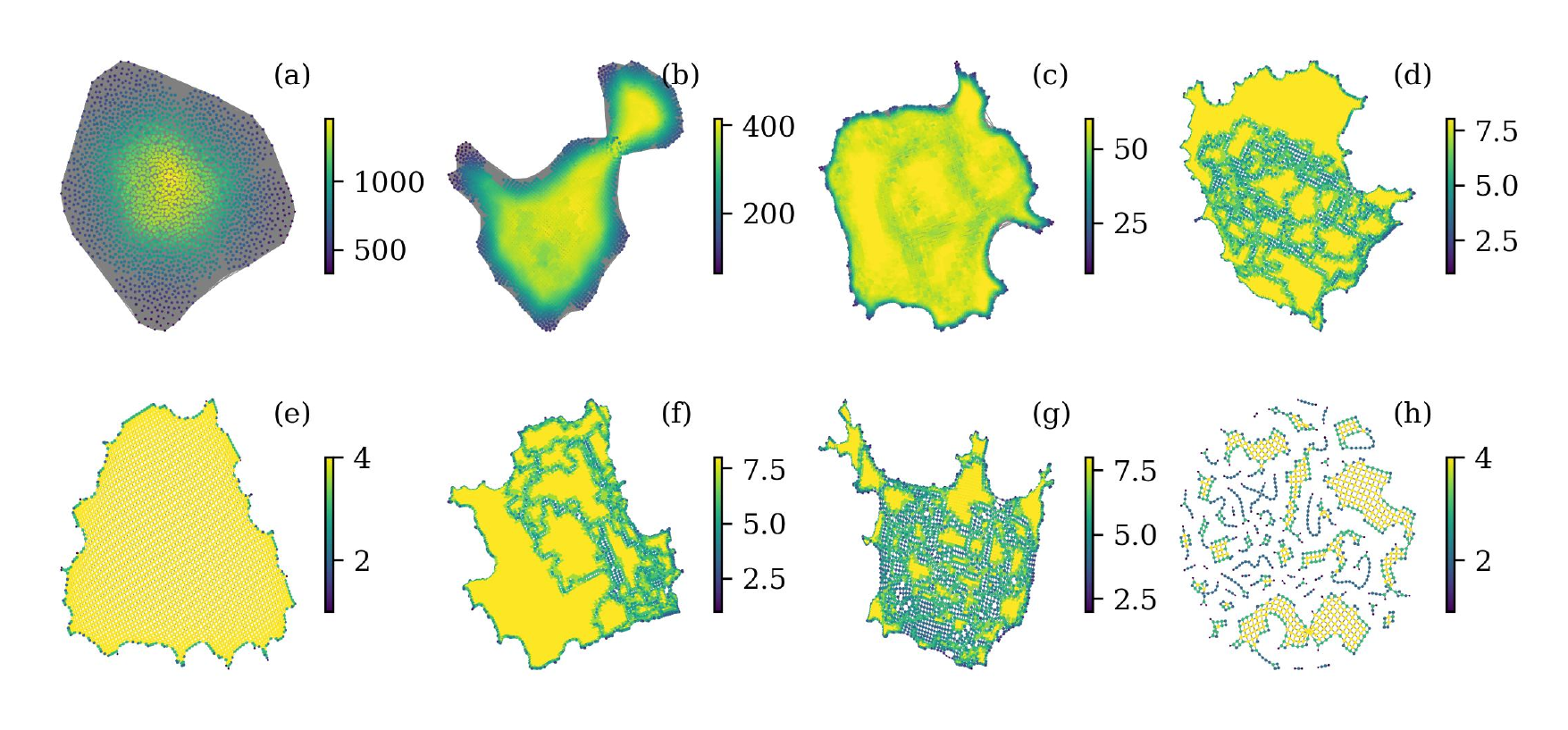}
    \caption{\label{fig:Network}Eight sample BTW networks with various $\mathcal{R}$, $N$, and $K_{\text{Avg}}$ values are shown in the figure. In these figures, Low-Active Streaks gradually emerge from left to right as $r_2$ increases in the first row, and as $r_1$ increases in the second row. (a) $\mathcal{R}=(0.1, 0.0)$, $N=2263$, $K_{Avg}=924$, (b) $\mathcal{R}=(0.1, 0.3)$, $N=4165$, $K_{Avg}=316$, (c) $\mathcal{R}=(0.1, 0.6)$, $N=7412$, $K_{Avg}=54$, (d) $\mathcal{R}=(0.1, 0.9)$, $N=6267$, $K_{Avg}=7$, $\mathcal{R}=(0.3, 0.9)$, $N=3165$, $K_{Avg}=4$, $\mathcal{R}=(0.5, 0.9)$, $N=7410$, $K_{Avg}=7$, $\mathcal{R}=(0.7, 0.9)$, $N=3816.$, $K_{Avg}=6$, $\mathcal{R}=(0.9, 0.9)$, $N=1087$, $K_{Avg}=3$.}
\end{figure*}

Examples of the network structure of $\mathcal{G}_{\mathcal{R}}(N)$ for variable $N$ (in the range $[1000, 9000]$) and variable $\mathcal{R}$ are shown in Fig.~\ref{fig:Network}. In the first row, $r_2$ ranges from $0$ to $0.9$ for a fixed $r_1 = 0.1$, while in the second row, $r_1$ ranges from $0.3$ to $0.9$ for a fixed $r_2 = 0.9$. The color bar represent the degree of the nodes. As shown in the first row, the network becomes increasingly sparse as $r_2$ increases, resulting in smaller degree centrality. For instance, the maximal degree in Fig.~\ref{fig:Network}a is approximately $1450$, whereas it decreases to $8$ for $\mathcal{R} = (0.1, 0.9)$ in Fig.~\ref{fig:Network}d. A similar trend is observed when $r_1$ increases, as evidenced in the second row, from left to right. An interesting feature of these graphs is the presence of well-separated communities: the nodes at the boundaries typically exhibit smaller degrees compared to those in the bulk. As an example, consider $\mathcal{R} = (0.9, 0.9)$ (Fig.~\ref{fig:Network}h), where the degrees are predominantly $4$ (for the bulk nodes), $3$ (for the boundary nodes), or $1$ (for weakly connected boundary nodes). Moving from left to right in Fig.~\ref{fig:Network} (first and second rows), we observe that instead of uniformly increasing the degree centralities, distinct structures begin to emerge. More specifically, a closer examination of the figures reveals the presence of less-active traces (streaks) formed by nodes with reduced activity, referred to as low-active streaks. For instance, in Fig.~\ref{fig:Network}f, among the highly active nodes (with degree $8$), there are traces corresponding to nodes with $k = 7$, which represent low-active streaks resulting from the sparsification process. In the first row, low-active streaks begin to form when $r_2$ is sufficiently large, while in the second row, a similar behavior is observed as $r_1$ increases.

To quantify the sparsification, we examined a variety of statistical observables, categorized into two groups: local measures and global measures. The local measures include the nodes' degree centrality, eigenvalue centrality, and clustering coefficient, while the global measures encompass betweenness centrality and closeness centrality, which will be described in detail in the subsequent sections.
\subsection{LOCAL MEASURES}~\label{SEC:local}
\subsubsection{Degree Centrality}~\label{SEC:degree}
\begin{figure*}[t]
    \centering   
    \includegraphics[width=\linewidth]{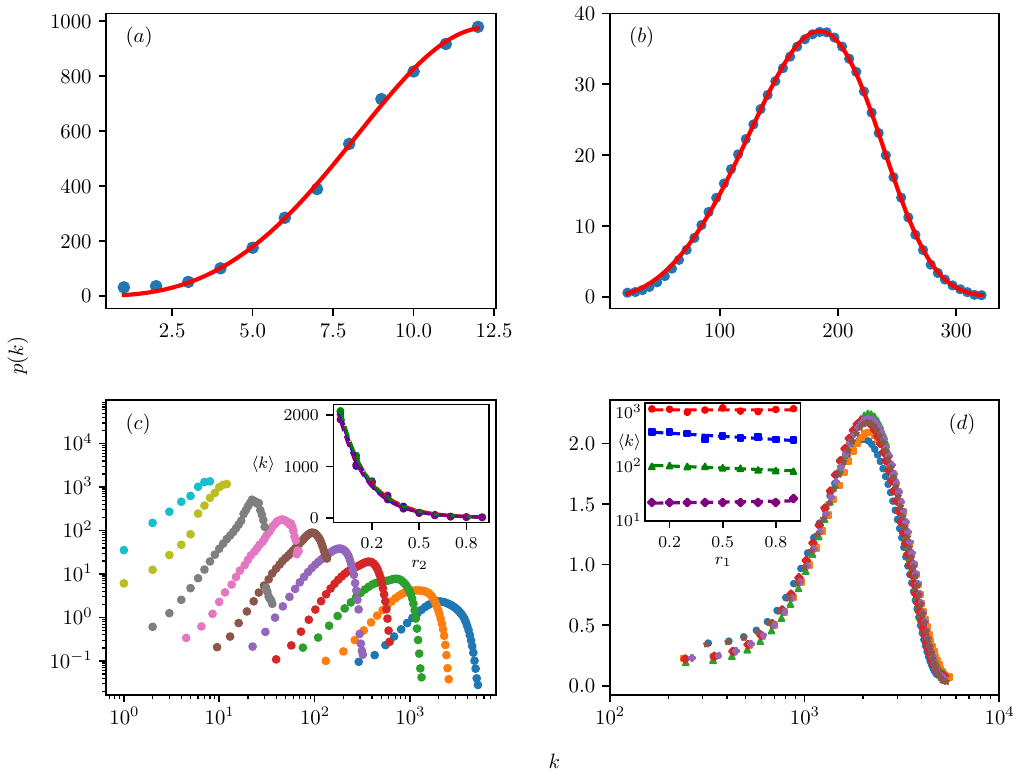}       
    \caption{The distribution function for the degree centrality $P(k)$. (a) represents $P(k)$ in the sparse phase $\mathcal{R}=(0.7,0.8)$ with a fit $G_{\alpha,\beta,\gamma}(k)$ with $\alpha=2.62\pm0.29$, $\beta=3.18\pm2.20$, and $\gamma=3.89\pm1.61$. Figure (b) is an example of the result for the dense phase $\mathcal{R}=(0.5,0.4)$ along with a fitting function $G_{\alpha,\beta,\gamma}(k)$ with $\alpha=2.29 \pm 0.02$, $\beta=2.94\pm 0.09$, and $\gamma=4.81\pm 0.06$. (c) and (d) show this function in terms of $r_1=0.5$ ($r_2=0$ to $0.9$) and $r_2=0.0$ ($r_1=0.1$ to $0.6$) respectively. In (c) $r_2$ decreases uniformly (with increment $0.1$) from left to right. In insets \( \textbf{c} \) and \( \textbf{d} \), the variables \( r_1 \) and \( r_2 \), respectively, vary from \( 0.1 \) to \( 0.7 \) with a step of \( 0.2 \), transitioning in symbol from circle to diamond.}
    \label{fig:degree}
\end{figure*}

The degree of nodes ($k$) in a complex network (CN) provides critical insights into node centrality, which can be interpreted as the frequency with which a specific node is utilized in the dynamics defined on the network. The degree of a node is defined as the total number of edges connected to it, representing the number of direct interactions that the node has. Degree centrality is widely regarded as one of the most important measures in CN analysis, serving as a fundamental metric that reflects the structural characteristics and connectivity of the network. The degree distribution function ($p(k)$) for various $\mathcal{R}$ values is presented in Fig.~\ref{fig:degree}, illustrating the sparse regime in (a) and the dense regime in (b). Additional degree distributions are shown for other $\mathcal{R}$ values: (c) as a function of $r_2$ and (d) as a function of $r_1$. Figure~\ref{fig:degree}c validates our sparsification arguments, demonstrating that as $r_2$ decreases, the peak of $p(k)$ shifts to the right, indicating higher $k$ values associated with the dense phase. Additionally, the graphs become broader with decreasing $r_2$, reflecting an increase in network inhomogeneity. These effects are less pronounced for $r_1$, as shown in Fig.~\ref{fig:degree}d, where the degree distribution graphs shift slightly to the right with decreasing $r_1$. The inset plots of diagrams (c) and (d) in Fig.~\ref{fig:degree} illustrate the variations of the peak position with respect to $r_2$ and $r_1$, respectively. As shown, the variations with $r_2$ follow an exponential trend, whereas those with $r_1$ exhibit a linear behavior. Although decreasing $r_1$ makes the network denser, the peak position remains largely unchanged. This distinction highlights the different roles of $r_1$ and $r_2$. The limited response of the average degree and the degree distribution function to changes in the similarity factor has already been observed in Fig.~\ref{fig:Network}. For instance, for $\mathcal{R}=(0.3,0.9)$, the average degree is $\bar{k}=4$, forming a compact, regular network where bulk sites connect to four neighbors, while boundary sites have $\bar{k}=3$. For $\mathcal{R}=(0.9,0.9)$, $\bar{k}=3$, with bulk sites maintaining the same connection pattern but an increased network sparseness leading to a higher proportion of boundary sites. As $r_1$ increases, low-activity streaks emerge, while the configuration of more-active nodes remains stable, resulting in minimal changes to the average degree $\bar{k}$.\\

To achieve the best fit for the degree distribution function, we employed a two-step approach. First, a genetic algorithm was used to determine the optimal initial parameters. Next, we refined the fit using the \textit{curve-fit} function from the \textit{scipy.optimize} library in Python, further enhancing the accuracy of the fitting process. Figure~\ref{fig:degree}(a, b) and the analysis across various $\mathcal{R}$ values reveal that $p(k)$ is well-approximated by a generalized Gamma distribution (GGD) function, described by
\begin{equation}
 G_{\alpha,\beta,\gamma}(k) = A(\alpha, \beta, \gamma) \cdot k^{\alpha} \cdot \exp\left[- \frac{1}{\beta} \cdot \left(\frac{k}{\bar{k}}\right)^{\gamma}\right]
    \label{eq:degre_fiting}
\end{equation}
where $\alpha$, $\beta$, and $\gamma$ are exponents that shape the function, and $A(\alpha, \beta, \gamma)$ is a normalization factor,
\begin{equation}
    A(\alpha, \beta, \gamma) = \frac{\gamma \cdot \beta^{-\frac{\alpha+1}{\gamma}}}{\bar{k}^{\alpha+1} \cdot \Gamma\left(\frac{\alpha+1}{\gamma}\right)}
\end{equation}
and $\Gamma(.)$ represents the Gamma function. The GGD is closely related to several other distributions, including the L\'evy stable distributions, generalized Gamma subordinator, and L\'evy processes (see Appendix \ref{SEC:appendices}). In the context of dynamical systems, the emergence of GGD in time series can be attributed to long-range correlations or memory effects, where the variable of interest at a given time is influenced by cumulative interactions or dependencies over multiple prior time steps. This dependency induces fluctuations that often follow a heavy-tailed distribution, frequently aligning with the behavior of GGD. However, in the context of complex networks, the appearance of GGD has been relatively rare. In self-organized critical systems, particularly fixed-energy sandpiles, both the avalanche size and the distribution of avalanche occurrence times have been shown to obey GGD \cite{manna2023nonstationary}. Additionally, GGD has been used to describe the probability density of earthquake recurrence times \cite{corral2004long}. Other time series, such as call duration in telecommunication systems, have also exhibited GGD behavior~\cite{Zonoozi_622908}. In directed Barabási-Albert and other local attachment networks, GGD has been utilized to calculate the minimum control count~\cite{Ruths2016}. Furthermore, it has been shown that in directed scientific, social, and scientific-social networks constructed from student relationships within certain scientific societies, the in-degree distribution follows GGD~\cite{Forsman_PhysRevSTPER.10.020122}. It is also instructive to examine the connection between GGD and power-law distributions for sufficiently small $\gamma$ and $k$ values:
\begin{eqnarray}
\label{Eq:Power-law}
&&\lim_{\gamma\ll 1}G_{\alpha,\beta,\gamma}(k)\approx A'(\alpha,\beta,\gamma)k^{\tau}, \ \ \ \\ 
&&A'(\alpha,\beta,\gamma)\equiv A(\alpha,\beta,\gamma)\left(e^{-1}\bar{k}^{\gamma}\right)^{\frac{1}{\beta}},\ \ \ \tau\equiv \alpha-\zeta,\ \ \zeta\equiv\frac{\gamma}{\beta}\nonumber,
\end{eqnarray} 
In other words, GGD transitions to a power-law distribution for sufficiently small $k$ values in the limit of small $\gamma$. Consequently, the behavior of the system in this regime represents a trade-off between $\alpha$ and $\gamma / \beta$. It is important to note that this conclusion holds for $k$ values satisfying the condition $\left(\frac{k}{\bar{k}}\right)^{\gamma} \ll e$. In the special case where $\gamma, \beta \to 0$ simultaneously such that $\zeta$ remains finite, the resulting exponent $\tau$ differs from $\alpha$.\\
\begin{figure*}[t]
    \centering       
    \includegraphics[width=\linewidth]{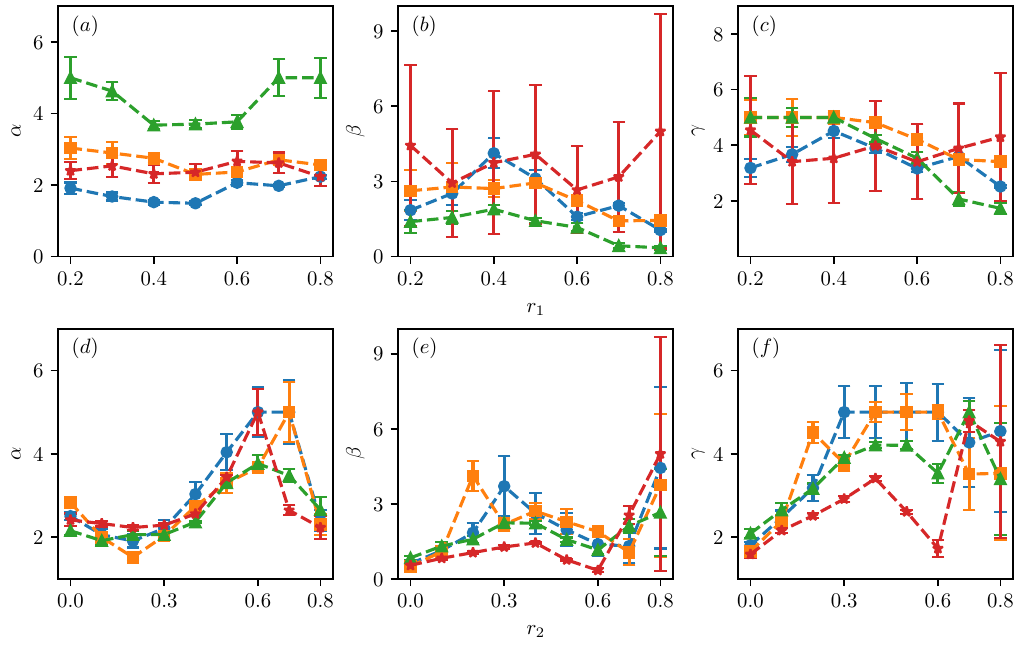}
    \caption{Variation of fitting exponents with $\mathcal{R}$: (a–c) Variation of \(\alpha\), \(\beta\), and \(\gamma\) with $r_1$. The circle, square, star, and triangle symbols in (a–c) correspond to \(r_2\) values of 0.2, 0.4, 0.6, and 0.8, respectively. (d–f) Variation of \(\alpha\), \(\beta\), and \(\gamma\) with $r_2$. The circle, square, star, and triangle symbols in (d–f) correspond to \(r_1\) values of 0.2, 0.4, 0.6, and 0.8, respectively.}
    \label{fig:fitt_parammter}
\end{figure*}
\begin{figure}
    \centering   
    \includegraphics{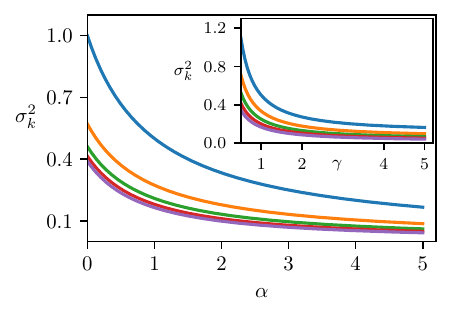}       
    \caption{The normalized variance $\sigma_k^2$ in terms of $\alpha$ (main) and $\gamma$ (inset). In the main part $\gamma=1,2,3,4$ and $5$, and in the inset $\alpha=1,2,3,4$ and $5$ from top to bottom.}
    \label{fig:variance}
\end{figure}
\begin{figure*}[t]
    \centering   
    \includegraphics[width=\linewidth]{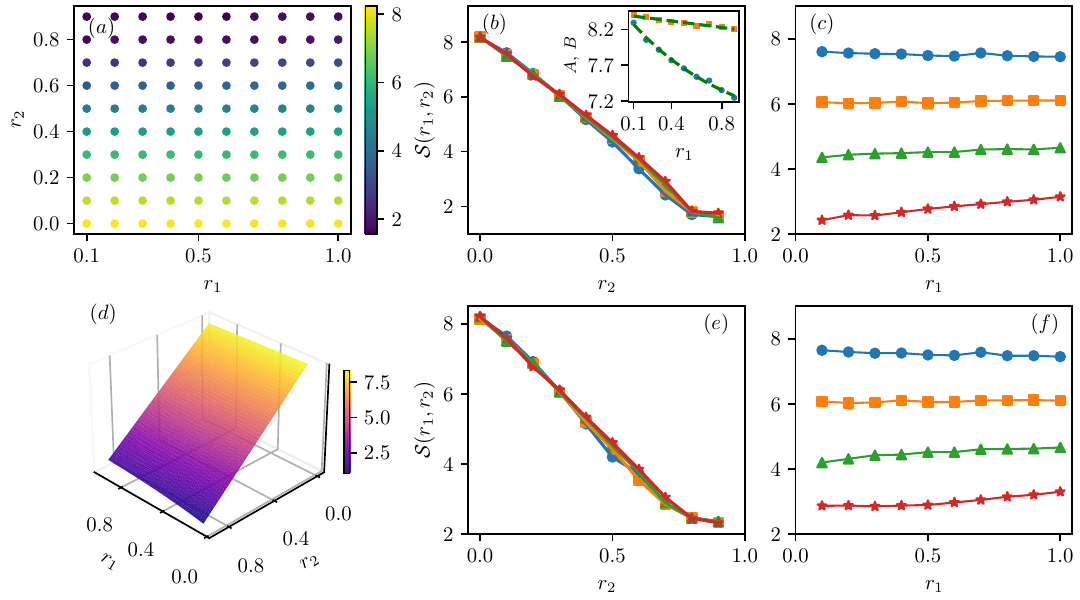}       
    \caption{Entropy: The plots (a) to (c) illustrate the variations of entropy as a function of $\mathcal {R}$ for the main data points of the network. Plot (d) represents the variations of entropy in a three-dimensional view. We observed that entropy depends on $r_1$ and $r_2$ according to the relationship $S = -A(r_1)r_2 + B(r_1)$, where $A$ is an exponential function of $r_1$ and $B$ is a linear function of $r_1$ (inset B). The plots (d) and (f) illustrate the variations of entropy as a function of $\mathcal {R}$ for the fitted function with fitting parameters in accordance with Eq \ref{Eq:shannon2}. In plots (b), (c), (e), and (f), the variables within the figure ($r_1$ or $r_2$) are represented by star, triangle, square, and circle symbols, corresponding to the values 0.1, 0.3, 0.5, and 0.7, respectively.}
    \label{fig:Entropy}
\end{figure*}
Figure~\ref{fig:fitt_parammter} depicts the exponents $\alpha$, $\beta$, and $\gamma$ as functions of $r_1$ and $r_2$. In Figs.~\ref{fig:fitt_parammter}a–c, $\alpha$, $\beta$, and $\gamma$ exhibit a subtle, non-monotonic dependence on $r_1$. Notably, for larger $r_1$ values—corresponding to sparser graphs—the error bar for $\beta$ increases, indicating a decline in the quality of the $\Gamma$ fitting. In contrast, the exponents show a stronger dependence on $r_2$, as illustrated in Figs.~\ref{fig:fitt_parammter}d–f. This behavior is consistent with the pronounced dependence of the average degree $\bar{k}$ on $r_2$ (see Fig.~\ref{fig:degree}). Specifically, Fig.~\ref{fig:fitt_parammter}d shows that as $r_2$ increases, $\alpha$ initially decreases for small $r_2$ values but then rises significantly within the range $r_2 \in [0.6-0.7]$ before returning to its initial value. The value of $\gamma$ exhibits a slight increase for small $r_2$ values and then either reverts to its initial value or stabilizes, depending on the $r_1$ value. A similar trend is observed for the $\beta$ exponent. It is worth noting that a smaller $\beta$ value corresponds to a faster decay of the GGD and indicates a less homogeneous network. Moreover, according to Eq.~\ref{Eq:Power-law}, a smaller $\gamma$ value results in a heavier-tailed GGD, approaching a power-law distribution in the $\gamma \to 0$ limit. Since the homogeneity of the graph is governed by the exponents $\alpha$, $\beta$, and $\gamma$, these results demonstrate how the parameter set $\mathcal{R}$ adjusts the network's structural characteristics. To further quantify this relationship, it is necessary to calculate the moments of $k$ as functions of the fitting parameters. In the continuum limit, where summations are transformed into integrals, the moments are expressed as:
\begin{eqnarray}
\label{Eq:expressions}
&&\gamma\left\langle \log k\right\rangle_{\text{GGD}}\rightarrow\psi^{(0)}\left(\frac{1+\alpha}{\gamma}\right)+\log\left(\beta\bar{k}^{\gamma}\right),\\
&&\left\langle k^{n}\right\rangle_{\text{GGD}}\rightarrow\beta^{\frac{n}{\gamma}}\bar{k}^{n}\frac{\Gamma[\frac{1+n+\alpha}{\gamma}]}{\Gamma[\frac{1+\alpha}{\gamma}]}\Rightarrow \frac{1}{\beta}\left\langle k^{\gamma}\right\rangle_{\text{GGD}}\rightarrow \bar{k}^{\gamma}\left(\frac{1+\alpha}{\gamma}\right),\nonumber
\end{eqnarray}
where $\psi^{(n)}(z)$ is the $n$th derivative of the digamma function $\psi(z)$. Using these expressions, one finds an analytical expression for the normalized variance of $k$:
\begin{equation}
\begin{split}
&\sigma_k^2\equiv \frac{\left\langle k^2\right\rangle_{\text{GGD}}-\left\langle k\right\rangle_{\text{GGD}}^2}{\left\langle k\right\rangle_{\text{GGD}}^2}=\frac{\Gamma\left(\frac{1+\alpha}{\gamma}\right)\Gamma\left(\frac{3+\alpha}{\gamma}\right)}{\Gamma\left(\frac{2+\alpha}{\gamma}\right)^2}-1,\\
& \frac{\left\langle k \right\rangle_{\text{GGD}} }{\bar{k}}=\frac{\beta^{\frac{1}{\gamma}}\Gamma\left(\frac{2+\alpha}{\gamma}\right)}{\Gamma\left(\frac{1+\alpha}{\gamma}\right)}.
\end{split} 
\end{equation}
We observe that the normalized variance is independent of $\beta$, whereas $\frac{\left\langle k \right\rangle_{\text{GGD}} }{\bar{k}}$ exhibits a power-law growth with $\beta$, with an exponent $1/\gamma$. Additionally, $\frac{\left\langle k \right\rangle_{\text{GGD}} }{\bar{k}}$ grows monotonically with $\gamma$ and decreases monotonically with $\alpha$. The normalized variance $\sigma_k^2$ is plotted in Fig.~\ref{fig:variance}, revealing that it is a monotonic decreasing function of both $\gamma$ and $\alpha$, while remaining independent of $\beta$. This indicates that the network becomes increasingly uniform as $\alpha$ and $\gamma$ grow. Furthermore, we note that $\sigma_k^2 \to \infty$ as $\gamma \to 0$, consistent with the fact that the network transitions to a scale-free regime in this limit, as predicted by Eq.~\ref{Eq:Power-law}. For power-law distributions, it is well-known that the variance diverges, which aligns with this observation.

\subsubsection{Shannon Entropy Associated with the Nodes' Degree}~\label{SEC:shannon}
In network theory, Shannon entropy is a widely used measure to quantify the diversity, randomness, or uncertainty associated with a specific property of the network~\cite{coon2018entropy}. To investigate the physical significance of the fitting parameters, we calculate the Shannon entropy associated with the degree distribution, defined as:
\begin{equation}
\mathcal{S}(\mathcal{R})\equiv -\sum_{k}P_{\mathcal{R}}(k)\log(P_{\mathcal{R}}(k)),
\label{Eq:shannon}
\end{equation}
where the summation over \(k\) runs over all available values of the degree. Shannon entropy provides insight into the variability or heterogeneity of node connectivity within the network. For regular (structured) networks, \(\mathcal{S}\) is low, whereas for networks with a wide variety of degrees, \(P(k)\) is more spread out, leading to higher entropy. This function is numerically calculated and presented in Fig.~\ref{fig:Entropy}. An analytical expression can be derived for systems where the degree centrality obeys the GGD. To achieve this, we substitute Eq.~\ref{eq:degre_fiting} into Eq.~\ref{Eq:shannon}, resulting in:
\begin{equation}
\begin{split}
\mathcal{S}^{\text{GGD}}_{\alpha,\beta,\gamma}=&-\log\left(A(\alpha, \beta, \gamma)\right)-\alpha\left\langle \log k\right\rangle_{\text{GGD}}\\
&+\frac{1}{\beta}\left\langle (k/\bar{k})^{\gamma}\right\rangle_{\text{GGD}},
\label{Eq:shannon2}
\end{split}
\end{equation}
where $\left\langle ... \right\rangle_{\text{GGD}}$ is the average with respect to GGD, i.e. for an observable $O_k$: 
\begin{equation}
\left\langle O_k \right\rangle_{\text{GGD}}\equiv \sum_{k}O_kG_{\alpha,\beta,\gamma}(k)\approx \frac{1}{N}\sum_{i=1}^N O_{k_i}.
\end{equation}
We see from Eq.~\ref{Eq:shannon2} that, taking apart the first term, $\mathcal{S}^{\text{GGD}}_{\alpha,\beta,\gamma}$ is comprised of two competing terms.  All in all, using the expressions Eqs.~\ref{Eq:expressions} we obtain
\begin{eqnarray}
\label{Eq:shannon3}  
\mathcal{S}^{\text{GGD}}_{\alpha,\beta,\gamma}&=&-\log\left(\frac{\gamma \cdot \beta^{-\frac{\alpha+1}{\gamma}}}{\bar{k}^{\alpha+1} \cdot \Gamma\left(\frac{\alpha+1}{\gamma}\right)}\right)\\
&&-\frac{\alpha}{\gamma}\left(\psi^{(0)}\left(\frac{1+\alpha}{\gamma}\right)
+\log\left(\beta\bar{k}^{\gamma}\right)\right)+\frac{1+\alpha}{\gamma},\nonumber
\end{eqnarray}
The results for numerical Shannon entropy are shown in Fig.~\ref{fig:Entropy}. In the Appnedix~\ref{App:Shannon} we plot the above equation in terms of $\alpha$, $\beta$, $\gamma$ and $\bar{k}$. The numerical Shannon entropy is reported in Fig.~\ref{fig:Entropy}a-c, while the Shannon entropy resulted from Eq.~\ref{Eq:shannon2} is represented in Fig.~\ref{fig:Entropy}d-f. We observe that $\mathcal{S}$ decreases almost linearly with $r_2$ the slope of which ($A$) and the width $B$ are plotted as functions of $r_1$ in the inset of Fig.~\ref{fig:Entropy}b. The dashed lines in this inset show the exponential decay for $A$ and linear decrease for $B$. Our numerical estimations reveal that the Shannon entropy fulfills the following relation
\begin{equation}
\mathcal{S}(r_1,r_2)=\mathcal{S}(0,0)-ar_2(1+b e^{-cr_1})-a'r_1,
\label{shannon4}
\end{equation}
where  $\mathcal{S}(0,0)=8.41\pm0.01$, $a=6.68\pm0.22$, $b=0.27\pm0.04$, $c=1.22\pm0.26$, and $a'=0.22\pm0.01$. A three-dimensional plot of this function is presented in Fig.~\ref{fig:Entropy} which shows a very good agreement with the numerical results. This gives, for the first time, a physical interpretation for $\mathcal{R}=(r_1,r_2)$, that is the set of parameters which tune the \textit{information content} of the network, i.e. the higher the amount of $r_1$ and $r_2$, the less informative the network is. The analytic form of the Shannon entropy in terms of $(\alpha,\beta,\gamma,N)$ is shown in Fig.~\ref{fig:Entropy_Analytical} in the appendix~\ref{App:Shannon}.
\subsubsection{Clustering Coefficient}~\label{SEC:cc}
The \textit{clustering coefficient} (\textit{cc}) is a measure that quantifies the tendency of nodes to form tightly-knit groups or clusters, providing insight into the local and global structure of the network, particularly the prevalence of triangles. As $\mathcal{R}$ (both components $r_1$ and $r_2$) increase, the network becomes sparser, and, as a result, one expects that the \textit{cc} decreases. The local \textit{cc} is defined as the ratio of the number of triangles involving the node to the number of possible triangles that could exist among its neighbors. Mathematically, the local \textit{cc} for the node $v$ is defined as:
\begin{equation}
cc(v)\equiv \frac{2n_{\bigtriangleup}(v)}{k_v(k_v-1)},
\label{Eq:ccCentrality}
\end{equation}
where $n_{\bigtriangleup}(v)$ is number of triangles involving $v$, and $k_v$ is the degree of the node $v$. A higher \textit{cc} indicates a greater tendency for nodes to form clusters or tightly-connected groups. It is used for identifying the small-world property of networks, where high clustering coexists with short average path lengths. \\

\begin{figure*}[t]
    \centering
{\includegraphics[width=\linewidth]{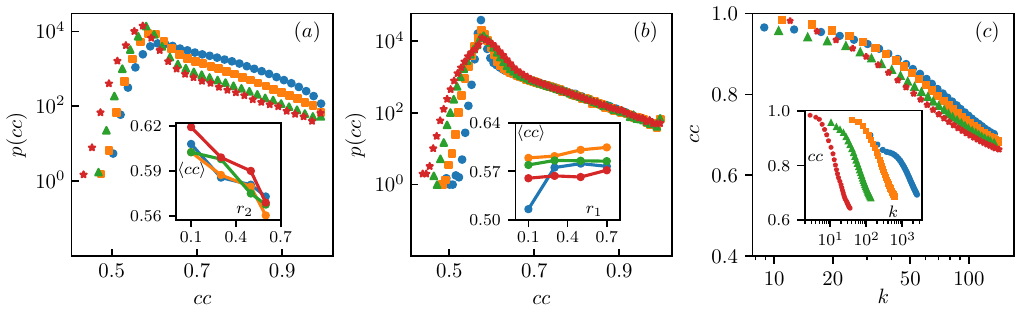}}
    \caption{\label{fig:clustring} Clustering coefficient distribution: (a main) clustering coefficient distribution for $\mathcal{R} = (0.5, r_2)$, where $r_2 = 0.1$, $0.3$, $0.5$, $0.6$ correspond to circle, Square, triangle and star symbols, respectively. (a inset) The peak points of the clustering coefficient for different \( r_1 \) values (0.2, 0.4, 0.5, 0.6) decrease as \( r_2 \) increases. (b main) clustering coefficient distribution for $\mathcal{R} = (r_1, 0.5)$, where $r_1 = 0.1$, $0.3$, $0.5$, $0.7$ correspond to circle, Square, triangle and star symbols, respectively. (b inset) the clustering coefficient peak points for different \( r_2 \) values (0.2, 0.4, 0.5, 0.6) increase as \( r_1 \) increases. (c main) Correlation between clustering coefficient and node degree for $\mathcal{R} = (r_1, 0.5)$, with $r_1 = 0.1$, $0.3$, $0.5$, $0.7$ as circle, square, triangle and star symbols. (c inset) Correlation for $\mathcal{R} = (0.5, r_2)$, with $r_2 = 0.1$, $0.3$, $0.5$, $0.7$ as circle, square, triangle and star symbols.}
\end{figure*}

Figure~\ref{fig:clustring} illustrates the distribution function of the \textit{clustering coefficient} (\textit{cc}) as a function of $\mathcal{R}$. A pronounced peak structure is observed, where the peak position $\left\langle cc \right\rangle$ decreases with increasing $r_2$ and increases with $r_1$. The decreasing trend of $\left\langle cc \right\rangle$ with $r_2$ can be attributed to the sparsification of the network as $r_2$ increases. This sparsification reduces the number of links, leading to a lower number of triangles, $n_{\bigtriangleup}$, and consequently smaller \textit{cc} values. As the network becomes sparser, the peak of the distribution progressively shifts towards lower \textit{cc} values. Interestingly, while increasing $r_1$ also dilutes the network, the \textit{cc} exhibits an opposite trend, increasing with larger $r_1$. This contrasting behavior underscores the distinct roles played by $r_1$ and $r_2$ in shaping the network's structural properties. This observation further highlights the distinct roles played by $r_1$ and $r_2$. According to Eq.~\ref{Eq:ccCentrality}, while both $n_{\bigtriangleup}$ (the number of triangles involving a node) and $k(k-1)$ (the total possible number of triangles) decrease with increasing $r_2$, the decrease in $n_{\bigtriangleup}$ is more pronounced, leading to an overall reduction in the clustering coefficient. In contrast, for increasing $r_1$, an inverse behavior is observed: the variation in $k(k-1)$ is relatively small (as shown in the inset of Fig.~\ref{fig:clustring}b), while $n_{\bigtriangleup}$ \textit{increases} with larger $r_1$. This implies that higher similarity among nodes leads to more connected configurations and higher clustering coefficients, even as the network becomes more diluted.\\

A negative correlation is observed between the clustering coefficient and the degree of the nodes, as illustrated in Fig.~\ref{fig:clustring}c. This indicates that while hub nodes are highly connected to other nodes, their neighbors are less likely to be interconnected. In preferential attachment models (not applicable here), hubs accumulate connections from diverse parts of the network, reducing the likelihood of direct connections among their neighbors. In our case, hubs correspond to highly active nodes whose neighbors have a lower probability of being directly connected to each other. For dense networks, this negative correlation is more pronounced, meaning that nodes with higher degrees typically exhibit lower clustering coefficients. As the network becomes sparser, the slope of this correlation decreases, yet the negative correlation remains evident. In other words, in sparser networks, high-degree nodes still tend to form fewer clusters, but this effect is less pronounced than in the dense case. The negative correlation between the clustering coefficient and the degree of nodes is often interpreted as evidence of a \textit{hierarchical structure}: the decrease in the clustering coefficient with increasing node degree reflects a layered network structure~\cite{nesterov2024clustering}. Small-scale clusters are predominantly formed by low-degree nodes, whereas high-degree nodes act as bridges between different sections of the network. These hub nodes exhibit fewer direct connections with similar nodes, resulting in a reduction in clustering among such nodes. Changes in the peak of the clustering coefficient distribution function with increasing parameter \( r_2 \) suggest a weakening of local clustering and a strengthening of hierarchical organization. This indicates that as the network transitions towards lower density, its hierarchical structure becomes more pronounced.
\subsubsection{Eigenvector Centrality}~\label{SEC:Eigen}
Eigenvector centrality is a measure used to quantify the influence of a node within a network. The centrality of each node is determined in proportion to the sum of the centralities of its connected neighbors. In simpler terms, nodes that are connected to highly central nodes will also exhibit high centrality. This metric is particularly valuable for identifying influential nodes by considering the importance of their neighbors. If the adjacency matrix of a network is denoted by \( A \equiv \{a_{ij}\}_{i,j=1}^N \), where \( a_{ij} = 1 \) if nodes \( i \) and \( j \) are interconnected, and \( a_{ij} = 0 \) otherwise, then the eigenvector centrality is defined through the eigenvector identity:
\begin{equation}
A\textbf{e}^{(\lambda)}=\lambda \textbf{e}^{(\lambda)},
\end{equation}
where $\lambda$ is an eigenvalue, and $\textbf{e}^{(\lambda)}=\left\lbrace e_i^{(\lambda)}\right\rbrace_{i=1}^N $ is the corresponding eigenvector. The eigenvector centrality is defined as $\textbf{e}\equiv \textbf{e}^{(\lambda_{\text{max}})}$, where $\lambda_{\text{max}}>0$ is the maximum eigenvalue. To see how the eigenvector centrality is related to the centrality of the neighbors, one may expand the eigenvalue relation as
\begin{equation}
e_i=\frac{1}{\lambda_{\text{max}}}\sum_j a_{ij}e_j=\frac{1}{\lambda_{\text{max}}}\sum_{j\in \mathcal{N}_i} e_j,
\end{equation}
where $\mathcal{N}_i$ is the set of neighbors of the node $i$, the cardinal of which is $|\mathcal{N}_i|=k_i$. This equation tells us that the centrality of the node $i$ is proportional to the sum of the centralities. For scale free networks the eigenvector centrality often exhibits localization, where most of the centrality is concentrated in the hubs and their immediate neighborhoods, and there is generally a strong, positive correlation (power-law) between the eigenvector centrality of a node and its degree. There is also a relation between the local \textit{fluctuation} of $e$ and the average degree of network, namely:
\begin{eqnarray}   
\delta e^2 &&\equiv \frac{1}{N\bar{e}^2}\sum_{i=1}^N\left(e_i-\bar{e}_{\delta_i}\right)^2\equiv\frac{1}{N\bar{e}^2}\sum_{i=1}^N\left(e_i-\frac{1}{k_i}\sum_{j\in \mathcal{N}_i} e_j\right)^2\nonumber\\
&&= \frac{1}{N\bar{e}^2}\sum_{i=1}^N\left(1-\frac{\lambda_m}{k_i}\right)^2e_i^2\approx \left(1-\frac{\lambda_m}{\bar{k}}\right)^2
\end{eqnarray}
where $\bar{e}_{\delta_i}\equiv\frac{1}{k_i}\sum_{j\in \mathcal{N}_i} e_j $, and $\bar{e}\equiv \frac{1}{N}\sum_{i=1}^Ne_i$. In the last line we replaced approximately $e_i$ by $\bar{e}$ and $k_i$ by $\bar{k}$, i.e. the averages. Therefore, we see that the local fluctuations of the eigenvector centrality is a function of the average degree centrality, and the largest eigenvalue. The above expression proves that for a network with equivalent nodes ($k_i=\bar{k}$ for all $i$s), $\delta e$ should be zero ($e_i=\bar{e}$ for all $i$s), and $\lambda_m=\bar{k}$.\\ 

\begin{figure*}[t]
  \centering      
    \includegraphics[width=\linewidth]{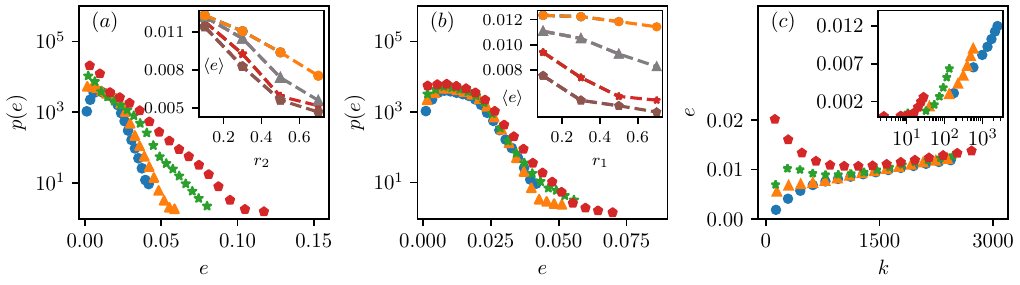}    
    \caption{\textbf{Eigenvector Centrality:} (a main) Eigenvector distribution for $\mathcal{R} = (0.1, r_2)$, where $r_2 = 0.1$, $0.3$, $0.5$, $0.7$ correspond to circle, triangle, star, and pentagon symbols, respectively. (a inset) Mean eigenvector value vs. $r_2$ for deferent $r_1$ values, where $r_1 = 0.1$, $0.3$, $0.5$, $0.7$ correspond to circle, triangle, star, and pentagon symbols, respectively. (b main) Eigenvector distribution for $\mathcal{R} = (r_1, 0.1)$, where $r_1 = 0.1$, $0.3$, $0.5$, $0.7$ correspond to circle, triangle, star, and pentagon. (b inset) Mean eigenvector value vs. $r_1$ for deferent $r_2$ values, where $r_2 = 0.1$, $0.3$, $0.5$, $0.7$ correspond to circle, triangle, star, and pentagon symbols, respectively. (c main) Correlation between eigenvector and node degree for $\mathcal{R} = (r_1, 0.1)$, with $r_1 = 0.1$, $0.3$, $0.5$, $0.7$ as circle, triangle, star, and pentagon. (c inset) Correlation for $\mathcal{R} = (0.1, r_2)$, with $r_2 = 0.1$, $0.3$, $0.5$, $0.7$ as circle, triangle, star, and pentagon.}
    \label{fig:eigenvector_centrality}
\end{figure*}

In our model, network sparsification significantly impacts the distribution of eigenvector centralities. The main effects can be summarized as follows:
\begin{itemize}
\item Sparsification tends to increase the number of disconnected or weakly connected components in the network. Nodes within these smaller or isolated components are typically weakly connected to the rest of the network, leading to lower eigenvector centralities. This results in an increase in the number of small eigenvalues.  
\item Sparsification often leaves behind a few highly connected sub-graphs or clusters. Nodes within these tightly interconnected sub-graphs tend to have high eigenvector centralities due to their strong mutual connections. As a result, there is a greater abundance of large eigenvalues, with centrality becoming concentrated within these dominant substructures.   
\item In dense networks, many nodes exhibit intermediate connectivity, meaning they are neither isolated nor part of dominant clusters. These nodes typically have intermediate eigenvector centralities. As the network becomes sparsified, many of these intermediate connections are removed, reducing the number of nodes with intermediate centralities. Consequently, the network exhibits fewer intermediate eigenvalues.
\end{itemize}
These structural transformations reflect the underlying shifts in centrality distribution upon sparsification of the network: increase in the population of nodes with small and large eigenvalues, while decrease for the intermediate eigenvalues. For small eigenvalues sparsification leads to more nodes become isolated or weakly connected, while for strongly connected nodes with large eigenvlaues denser clusters result, which leads to a shift to higher eigenvalues. The inverse is expected for intermediate eigenvalues: these nodes either lose connections, lowering their eigenvalues, or merge into highly connected clusters, raising their eigenvalues. This results in a reduction in the number of intermediate eigenvalues, and an increase in the abundance of nodes with small and high enough $e$ values. The distribution function of eigenvector centrality $p(e)$ is shown in Fig.~\ref{fig:eigenvector_centrality}. Figure~\ref{fig:eigenvector_centrality}a shows $p(e)$ for variable $r_2$ (inset shows the average in terms of $r_2$ for various $r_1$ values), while Fig.~\ref{fig:eigenvector_centrality}b shows the same with $r_2\leftrightarrows r_1$. Again, we see that the response of the eigenvalue centrality to the change of $r_2$ is much larger than $r_1$.\\

The correlation between eigenvector centrality ($e$) and degree ($k$) is shown in Fig.~\ref{fig:eigenvector_centrality}. We observe that some eigenvector centralities have positive values while others are negative, reflecting the varying influence of different clusters on the broader network structure. A negative correlation between $e$ and $k$ indicates that high-degree nodes are connected to low-centrality (peripheral) nodes. Figure~\ref{fig:eigenvector_centrality}c (main) illustrates that, for high $r_1$ values and low $r_2$ values, the correlation becomes negative in a certain range of $k$. For example, in the case of $\mathcal{R}=(0.7,0.1)$ and for $k \lesssim 1500$, a negative correlation is observed. In these situations, high-degree nodes tend to be disconnected from the core of the network, where central nodes reinforce each other's centrality. This behavior is associated with anomalous trends in other network measures, as will be discussed in the following sections.
\subsection{GLOBAL MEASURES}~\label{SEC:global}
We evaluate the importance of nodes within the network by analyzing two key global centrality measures: \textit{closeness centrality} and \textit{betweenness centrality}. Each of these measures offers a unique perspective on the role of nodes in terms of influence, connectivity, and their position within the overall network structure. By examining and comparing these centrality measures, we can gain a comprehensive understanding of the network's topological characteristics and the significance of its nodes.

\subsubsection{Closeness Centrality}~\label{SEC:closeness}
The \textit{closeness centrality} is inversely related to the average distance from a typical node to all other vertices in the network~\cite{newman2018networks}. It can be compared to the inverse of the average shortest path length in the network. The normalized closeness centrality of node $i$ is defined as
\begin{equation}
c_i^{-1}\equiv \frac{1}{N-1}\sum_{j=1}^N d(i,j),
\end{equation}
where $d(i,j)$ is the shortest path distance (geodesic distance) between two nodes $i$ and $j$. A node with high closeness centrality is, on average, closer to all other nodes and can disseminate information more efficiently. In a dense network, where many nodes are close to each other, the closeness centrality tends to be high, approaching 1. In contrast, for sparser networks, the closeness centrality is expected to be lower.\\
\begin{figure*}[t]
  \centering
    {\includegraphics[width=\linewidth]{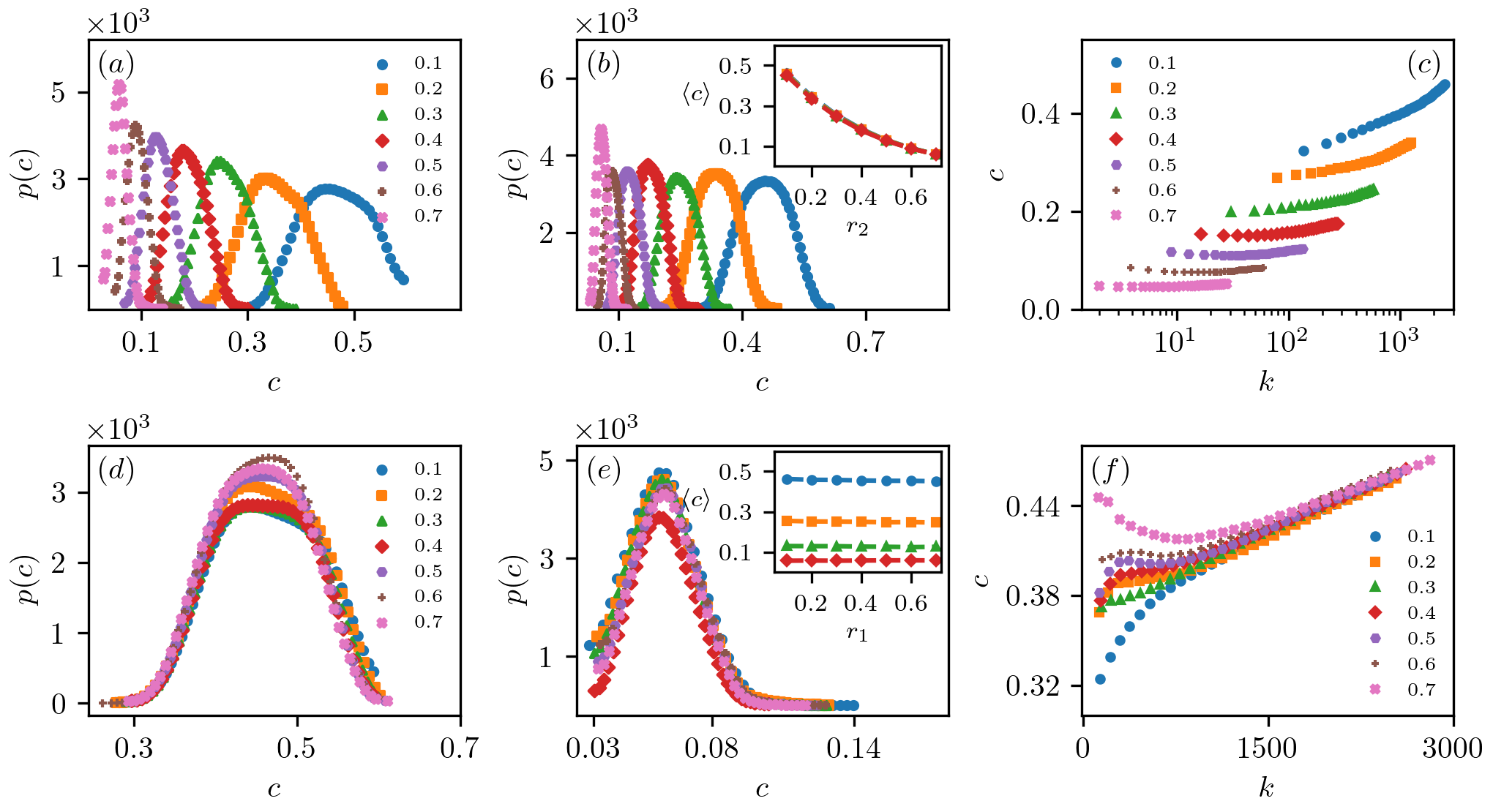}}    
    \caption{\textbf{Closeness Centrality:} (a) The closeness distribution for $\mathcal{R} = (0.1, r_2)$, where $r_2$ takes values from $0.1$ to $0.7$, corresponding to symbols from circle to cross, respectively. (b main) The closeness distribution for $\mathcal{R} = (0.7, r_2)$, where $r_2$ takes values from $0.1$ to $0.7$, corresponding to symbols from circle to cross, respectively. (b inset) The mean closeness as a function of $r_2$ for different $r_1$ values of $0.1$, $0.3$, $0.5$, and $0.7$, represented by circle, square, triangle, and diamond symbols, respectively. (c) The correlation between closeness and node degree for $r_1 = 0.1$ and $r_2$ varying from $0.1$ to $0.7$, represented by symbols from circle to cross, respectively. (d) The closeness distribution for $\mathcal{R} = (r_1, 0.1)$, where $r_1$ takes values from $0.1$ to $0.7$, corresponding to symbols from circle to cross, respectively. (e main) The closeness distribution for $\mathcal{R} = (r_1, 0.7)$, where $r_1$ takes values from $0.1$ to $0.7$, corresponding to symbols from circle to cross, respectively. (e inset) The mean closeness as a function of $r_1$ for different $r_12$ values of $0.1$, $0.3$, $0.5$, and $0.7$, represented by circle, square, triangle, and diamond symbols, respectively. (f) The correlation between closeness and node degree for $r_2 = 0.1$ and $r_1$ varying from $0.1$ to $0.7$, represented by symbols from circle to cross, respectively.}
    \label{fig:closeness_centrality}
\end{figure*}
As seen in Fig.~\ref{fig:closeness_centrality}, the distribution function exhibits a peaked structure, with the peak value $c$ decreasing as $r_2$ increases. This is consistent with the expectation that increasing $r_2$ dilutes the network. As $r_2$ decreases, the closeness centrality increases, the distribution broadens, and the fluctuations of $c$ increase, leading to a more heterogeneous network. These trends are consistent with the observations for the degree distribution function (see Fig.~\ref{fig:degree}c). The dependence of the peak value and the distribution width on $r_1$ is much weaker, as seen in Figs.~\ref{fig:closeness_centrality}d and~\ref{fig:closeness_centrality}e. This can be understood by noting that $r_1$ does not significantly alter the density (or sparsity) of the network. However, it is worth noting that $r_1$ slightly affects the width of $p(c)$.\\

When examining the relationship between degree and closeness centrality across different values of $r_2$, a consistently positive correlation between these two metrics is observed. However, as the parameter $r_2$ increases, this positive correlation gradually weakens. Interestingly, at some value of $r_1$, the correlation not only diminishes but actually becomes negative (Fig. \ref{fig:closeness_centrality}f). This shift indicates that increasing $r_1$ significantly alters the structural dynamics of the network, leading to an inverse correlation between degree and closeness centrality. Remarkably, we observe that this negative correlation occurs precisely where the $e-k$ correlation becomes negative (high $r_1$, low $r_2$, and low $k$ values). The observation of a negative correlation between eigenvector centrality and degree centrality, as well as between closeness and degree in a limited part of the degree spectrum, is intriguing and suggests nuanced structural features. This phenomenon indicates that nodes with higher degrees are poorly connected to other central nodes. These high-degree nodes are not part of the network’s core structure, where central nodes reinforce each other’s centrality. Such a scenario could occur in dis-assortative networks (where high-degree nodes connect more often to low-degree nodes) or in the presence of a structural bottleneck that prevents these nodes from being central in the eigenvector sense. In this regime, the network is expected to exhibit a community structure.

\subsubsection{Betweenness Centrality}~\label{SEC:betweenness}
The betweenness centrality is a global measure that quantifies the way nodes are placed directly "between" other nodes, i.e., they are in the shortest path between other nodes. Mathematically, the betweenness centrality of node $i$ is defined as
\begin{equation}
b_i=\frac{\sum_{j_1j_2}\sigma_{j_1ij_2}}{\sum_{j_1j_2}\sigma_{j_1j_2}},
\end{equation}
$\sigma_{j_1ij_2}$ represents the number of shortest paths between nodes $j_1$ and $j_2$ that pass through node $i$, and $\sigma_{j_1j_2}$ is the total number of shortest paths between $j_1$ and $j_2$. Betweenness centrality plays a critical role in maintaining network stability and facilitating information flow.\\
\begin{figure*}[t]
  \centering
    {\includegraphics[width=\linewidth]{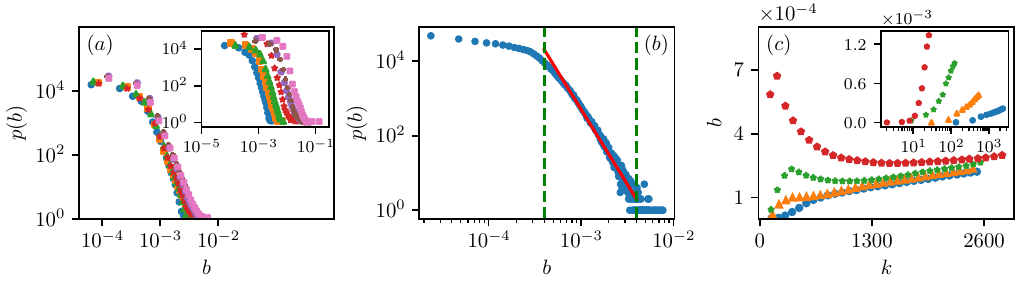}}\,
    \caption{\textbf{Betweenness Centrality:} (a main) The Betweenness distribution for $\mathcal{R} = (r_1, 0.1)$, where $r_1$ takes values from $0.1$ to $0.7$, represented by symbols from circle to cross. (ainset) The Betweenness distribution for $\mathcal{R} = (0.1, r_2)$, where $r_2$ takes values from $0.1$ to $0.7$, represented by symbols from circle to cross. (b) $\mathcal{R} = (0.7, 0.1)$: The slope in the selected part is -4.02. (c main) Correlation between Betweenness and node degree for $\mathcal{R} = (r_1, 0.1)$, with $r_1 = 0.1$, $0.3$, $0.5$, $0.7$ as circle, triangle, star, and pentagon. (c inset) Correlation for $\mathcal{R} = (0.1, r_2)$, with $r_2 = 0.1$, $0.3$, $0.5$, $0.7$ as circle, triangle, star, and pentagon.}

    \label{fig:Betweenness_centrality}
\end{figure*}
The results are depicted in Fig.~\ref{fig:Betweenness_centrality}. While the other measures exhibit peaked distribution functions, it is interesting to note that the distribution function of the betweenness centrality, $p(b)$, is endowed with a fat tail, showing a power-law form in certain limits and ranges of $b$. For instance, specifically for $\mathcal{R}=(0.7,0.1)$, the power-law distribution is observed from $2 \times 10^{-4}$ to $3 \times 10^{-3}$ (Fig.~\ref{fig:Betweenness_centrality}b). Notably, within these ranges, the exponents of the power-law distributions remain relatively consistent (Fig.~\ref{fig:Betweenness_centrality}a). As the network is diluted (i.e., by increasing $r_1$ and $r_2$), the distribution functions shift: once again, the dependence on $r_2$ is stronger than on $r_1$. The correlation between degree and betweenness centrality is shown in Fig.~\ref{fig:Betweenness_centrality}c. Interestingly, a negative correlation is again observed in the same region where the $c-k$ and $e-k$ correlations are negative.
\section{Concluding remarks}~\label{SEC:concl}
This paper tackles a longstanding challenge in complex systems: mapping sandpiles to complex networks, a framework that sheds light on the hidden properties and dynamics of the original system. The methods we propose are not only effective for this mapping but also hold potential for application to other SOC systems characterized by avalanche dynamics. The mapping relies on two control parameters, $\mathcal{R}=(r_1, r_2)$, which fundamentally shape the resulting network. Specifically, $r_1$ captures the similarity between local activities, while $r_2$ sets the threshold for converting a weighted network into a binary one. Our results demonstrate that network density is primarily influenced by $r_2$, with $r_1$ playing a subtler role by tuning the network's geometry. As $r_1$ and $r_2$ increase, the network undergoes a transition from a dense to a sparse phase, accompanied by the emergence of low-activity streaks. We found that the degree distribution $k$ follows a generalized Gamma distribution (GGD), providing an analytical framework for studying the system. Using the GGD, we derived an expression for the Shannon entropy, which aligns well with numerical findings. This analytic form captures the dependency of Shannon entropy on $\mathcal{R}$.\\

To explore the network's structure, we examined both local and global measures. Local measures, such as the clustering coefficient ($cc$) and eigenvector centrality ($e$), exhibited peaked distributions, while global measures, including closeness centrality ($c$) and betweenness centrality ($b$), revealed distinct patterns. Notably, the distribution of $b$ followed a power-law within a finite range. Interestingly, we observed negative correlations between $e-k$, $c-k$, and $b-k$ in specific parameter regimes, signaling anomalous network behavior. These correlations indicate that high-degree nodes can be poorly connected to other central nodes, leading to a non-trivial community structure. This phenomenon becomes prominent for large $r_1$ and small $r_2$ values, where global minima in $e$, $c$, and $b$ emerge, as observed in the correlation graphs. In our model, the non-trivial community structure arises due to the formation of low-activity streaks, highlighting a complex interplay between local activity similarity and global thresholds. These findings offer valuable insights into the organization of complex networks derived from SOC systems and open avenues for further exploration of their structural and dynamical properties.

\bibliography{refs}

\begin{thebibliography}{64}%
\makeatletter
\providecommand \@ifxundefined [1]{%
 \@ifx{#1\undefined}
}%
\providecommand \@ifnum [1]{%
 \ifnum #1\expandafter \@firstoftwo
 \else \expandafter \@secondoftwo
 \fi
}%
\providecommand \@ifx [1]{%
 \ifx #1\expandafter \@firstoftwo
 \else \expandafter \@secondoftwo
 \fi
}%
\providecommand \natexlab [1]{#1}%
\providecommand \enquote  [1]{``#1''}%
\providecommand \bibnamefont  [1]{#1}%
\providecommand \bibfnamefont [1]{#1}%
\providecommand \citenamefont [1]{#1}%
\providecommand \href@noop [0]{\@secondoftwo}%
\providecommand \href [0]{\begingroup \@sanitize@url \@href}%
\providecommand \@href[1]{\@@startlink{#1}\@@href}%
\providecommand \@@href[1]{\endgroup#1\@@endlink}%
\providecommand \@sanitize@url [0]{\catcode `\\12\catcode `\$12\catcode
  `\&12\catcode `\#12\catcode `\^12\catcode `\_12\catcode `\%12\relax}%
\providecommand \@@startlink[1]{}%
\providecommand \@@endlink[0]{}%
\providecommand \url  [0]{\begingroup\@sanitize@url \@url }%
\providecommand \@url [1]{\endgroup\@href {#1}{\urlprefix }}%
\providecommand \urlprefix  [0]{URL }%
\providecommand \Eprint [0]{\href }%
\providecommand \doibase [0]{https://doi.org/}%
\providecommand \selectlanguage [0]{\@gobble}%
\providecommand \bibinfo  [0]{\@secondoftwo}%
\providecommand \bibfield  [0]{\@secondoftwo}%
\providecommand \translation [1]{[#1]}%
\providecommand \BibitemOpen [0]{}%
\providecommand \bibitemStop [0]{}%
\providecommand \bibitemNoStop [0]{.\EOS\space}%
\providecommand \EOS [0]{\spacefactor3000\relax}%
\providecommand \BibitemShut  [1]{\csname bibitem#1\endcsname}%
\let\auto@bib@innerbib\@empty
\bibitem [{\citenamefont {Bak}\ \emph {et~al.}(1987)\citenamefont {Bak},
  \citenamefont {Tang},\ and\ \citenamefont {Wiesenfeld}}]{bak1987self}%
  \BibitemOpen
  \bibfield  {author} {\bibinfo {author} {\bibfnamefont {P.}~\bibnamefont
  {Bak}}, \bibinfo {author} {\bibfnamefont {C.}~\bibnamefont {Tang}},\ and\
  \bibinfo {author} {\bibfnamefont {K.}~\bibnamefont {Wiesenfeld}},\ }\bibfield
   {title} {\bibinfo {title} {Self-organized criticality: An explanation of the
  1/f noise},\ }\href {https://doi.org/10.1103/PhysRevLett.59.381} {\bibfield
  {journal} {\bibinfo  {journal} {Physical review letters}\ }\textbf {\bibinfo
  {volume} {59}},\ \bibinfo {pages} {381} (\bibinfo {year} {1987})}\BibitemShut
  {NoStop}%
\bibitem [{\citenamefont {Kadanoff}\ \emph {et~al.}(1989)\citenamefont
  {Kadanoff}, \citenamefont {Nagel}, \citenamefont {Wu},\ and\ \citenamefont
  {Zhou}}]{kadanoff1989scaling}%
  \BibitemOpen
  \bibfield  {author} {\bibinfo {author} {\bibfnamefont {L.~P.}\ \bibnamefont
  {Kadanoff}}, \bibinfo {author} {\bibfnamefont {S.~R.}\ \bibnamefont {Nagel}},
  \bibinfo {author} {\bibfnamefont {L.}~\bibnamefont {Wu}},\ and\ \bibinfo
  {author} {\bibfnamefont {S.-m.}\ \bibnamefont {Zhou}},\ }\bibfield  {title}
  {\bibinfo {title} {Scaling and universality in avalanches},\ }\href
  {https://doi.org/10.1103/PhysRevA.39.6524} {\bibfield  {journal} {\bibinfo
  {journal} {Physical Review A}\ }\textbf {\bibinfo {volume} {39}},\ \bibinfo
  {pages} {6524} (\bibinfo {year} {1989})}\BibitemShut {NoStop}%
\bibitem [{\citenamefont {Jensen}\ \emph {et~al.}(1989)\citenamefont {Jensen},
  \citenamefont {Christensen},\ and\ \citenamefont {Fogedby}}]{Jensen_1989}%
  \BibitemOpen
  \bibfield  {author} {\bibinfo {author} {\bibfnamefont {H.~J.}\ \bibnamefont
  {Jensen}}, \bibinfo {author} {\bibfnamefont {K.}~\bibnamefont
  {Christensen}},\ and\ \bibinfo {author} {\bibfnamefont {H.~C.}\ \bibnamefont
  {Fogedby}},\ }\bibfield  {title} {\bibinfo {title} {1/f noise, distribution
  of lifetimes, and a pile of sand},\ }\href
  {https://doi.org/10.1103/PhysRevB.40.7425} {\bibfield  {journal} {\bibinfo
  {journal} {Phys. Rev. B}\ }\textbf {\bibinfo {volume} {40}},\ \bibinfo
  {pages} {7425} (\bibinfo {year} {1989})}\BibitemShut {NoStop}%
\bibitem [{\citenamefont {Dhar}(1990)}]{dhar1990self}%
  \BibitemOpen
  \bibfield  {author} {\bibinfo {author} {\bibfnamefont {D.}~\bibnamefont
  {Dhar}},\ }\bibfield  {title} {\bibinfo {title} {Self-organized critical
  state of sandpile automaton models},\ }\href
  {https://doi.org/10.1103/PhysRevLett.64.1613} {\bibfield  {journal} {\bibinfo
   {journal} {Physical Review Letters}\ }\textbf {\bibinfo {volume} {64}},\
  \bibinfo {pages} {1613} (\bibinfo {year} {1990})}\BibitemShut {NoStop}%
\bibitem [{\citenamefont {Najafi}\ \emph {et~al.}(2012)\citenamefont {Najafi},
  \citenamefont {Moghimi-Araghi},\ and\ \citenamefont
  {Rouhani}}]{Najafi2012Moghimi}%
  \BibitemOpen
  \bibfield  {author} {\bibinfo {author} {\bibfnamefont {M.~N.}\ \bibnamefont
  {Najafi}}, \bibinfo {author} {\bibfnamefont {S.}~\bibnamefont
  {Moghimi-Araghi}},\ and\ \bibinfo {author} {\bibfnamefont {S.}~\bibnamefont
  {Rouhani}},\ }\bibfield  {title} {\bibinfo {title} {Avalanche frontiers in
  the dissipative abelian sandpile model and off-critical schramm-loewner
  evolution},\ }\href {https://doi.org/10.1103/PhysRevE.85.051104} {\bibfield
  {journal} {\bibinfo  {journal} {Phys. Rev. E}\ }\textbf {\bibinfo {volume}
  {85}},\ \bibinfo {pages} {051104} (\bibinfo {year} {2012})}\BibitemShut
  {NoStop}%
\bibitem [{\citenamefont {Manna}(1999)}]{manna1999sandpile}%
  \BibitemOpen
  \bibfield  {author} {\bibinfo {author} {\bibfnamefont {S.}~\bibnamefont
  {Manna}},\ }\bibfield  {title} {\bibinfo {title} {Sandpile models of
  self-organized criticality},\ }\href {https://doi.org/here} {\bibfield
  {journal} {\bibinfo  {journal} {Current Science}\ ,\ \bibinfo {pages} {388}}
  (\bibinfo {year} {1999})}\BibitemShut {NoStop}%
\bibitem [{\citenamefont {Bak}\ \emph {et~al.}(1988)\citenamefont {Bak},
  \citenamefont {Tang},\ and\ \citenamefont {Wiesenfeld}}]{bak1988self}%
  \BibitemOpen
  \bibfield  {author} {\bibinfo {author} {\bibfnamefont {P.}~\bibnamefont
  {Bak}}, \bibinfo {author} {\bibfnamefont {C.}~\bibnamefont {Tang}},\ and\
  \bibinfo {author} {\bibfnamefont {K.}~\bibnamefont {Wiesenfeld}},\ }\bibfield
   {title} {\bibinfo {title} {Self-organized criticality},\ }\href
  {https://doi.org/10.1103/PhysRevA.38.364} {\bibfield  {journal} {\bibinfo
  {journal} {Physical review A}\ }\textbf {\bibinfo {volume} {38}},\ \bibinfo
  {pages} {364} (\bibinfo {year} {1988})}\BibitemShut {NoStop}%
\bibitem [{\citenamefont {Tang}\ and\ \citenamefont
  {Bak}(1988)}]{tang1988critical}%
  \BibitemOpen
  \bibfield  {author} {\bibinfo {author} {\bibfnamefont {C.}~\bibnamefont
  {Tang}}\ and\ \bibinfo {author} {\bibfnamefont {P.}~\bibnamefont {Bak}},\
  }\bibfield  {title} {\bibinfo {title} {Critical exponents and scaling
  relations for self-organized critical phenomena},\ }\href
  {https://doi.org/10.1103/PhysRevLett.60.2347} {\bibfield  {journal} {\bibinfo
   {journal} {Physical Review Letters}\ }\textbf {\bibinfo {volume} {60}},\
  \bibinfo {pages} {2347} (\bibinfo {year} {1988})}\BibitemShut {NoStop}%
\bibitem [{\citenamefont {Redig}\ \emph {et~al.}(2012)\citenamefont {Redig},
  \citenamefont {Ruszel},\ and\ \citenamefont {Saada}}]{redig2012abelian}%
  \BibitemOpen
  \bibfield  {author} {\bibinfo {author} {\bibfnamefont {F.}~\bibnamefont
  {Redig}}, \bibinfo {author} {\bibfnamefont {W.}~\bibnamefont {Ruszel}},\ and\
  \bibinfo {author} {\bibfnamefont {E.}~\bibnamefont {Saada}},\ }\bibfield
  {title} {\bibinfo {title} {The abelian sandpile model on a random binary
  tree},\ }\href {https://doi.org/10.1007/s10955-012-0498-6} {\bibfield
  {journal} {\bibinfo  {journal} {Journal of Statistical Physics}\ }\textbf
  {\bibinfo {volume} {147}},\ \bibinfo {pages} {653} (\bibinfo {year}
  {2012})}\BibitemShut {NoStop}%
\bibitem [{\citenamefont {Welsh}\ and\ \citenamefont
  {Merino}(2000)}]{welsh2000potts}%
  \BibitemOpen
  \bibfield  {author} {\bibinfo {author} {\bibfnamefont {D.~J.}\ \bibnamefont
  {Welsh}}\ and\ \bibinfo {author} {\bibfnamefont {C.}~\bibnamefont {Merino}},\
  }\bibfield  {title} {\bibinfo {title} {The potts model and the tutte
  polynomial},\ }\href {https://doi.org/0.1063/1.533181} {\bibfield  {journal}
  {\bibinfo  {journal} {Journal of Mathematical Physics}\ }\textbf {\bibinfo
  {volume} {41}},\ \bibinfo {pages} {1127} (\bibinfo {year}
  {2000})}\BibitemShut {NoStop}%
\bibitem [{\citenamefont {Pastor-Satorras}\ and\ \citenamefont
  {Vespignani}(2000)}]{pastor2000universality}%
  \BibitemOpen
  \bibfield  {author} {\bibinfo {author} {\bibfnamefont {R.}~\bibnamefont
  {Pastor-Satorras}}\ and\ \bibinfo {author} {\bibfnamefont {A.}~\bibnamefont
  {Vespignani}},\ }\bibfield  {title} {\bibinfo {title} {Universality classes
  in directed sandpile models},\ }\href
  {https://doi.org/10.1088/0305-4470/33/3/101} {\bibfield  {journal} {\bibinfo
  {journal} {Journal of Physics A: Mathematical and General}\ }\textbf
  {\bibinfo {volume} {33}},\ \bibinfo {pages} {L33} (\bibinfo {year}
  {2000})}\BibitemShut {NoStop}%
\bibitem [{\citenamefont {De~Luca}\ and\ \citenamefont
  {Franchini}(2013)}]{De_Luca_2013}%
  \BibitemOpen
  \bibfield  {author} {\bibinfo {author} {\bibfnamefont {A.}~\bibnamefont
  {De~Luca}}\ and\ \bibinfo {author} {\bibfnamefont {F.}~\bibnamefont
  {Franchini}},\ }\bibfield  {title} {\bibinfo {title} {Approaching the
  restricted solid-on-solid critical points through entanglement: One model for
  many universalities},\ }\bibfield  {journal} {\bibinfo  {journal} {Physical
  Review B}\ }\textbf {\bibinfo {volume} {87}},\ \href
  {https://doi.org/10.1103/physrevb.87.045118} {10.1103/physrevb.87.045118}
  (\bibinfo {year} {2013})\BibitemShut {NoStop}%
\bibitem [{\citenamefont {Ali}\ and\ \citenamefont
  {Dhar}(1995)}]{ali1995structure}%
  \BibitemOpen
  \bibfield  {author} {\bibinfo {author} {\bibfnamefont {A.~A.}\ \bibnamefont
  {Ali}}\ and\ \bibinfo {author} {\bibfnamefont {D.}~\bibnamefont {Dhar}},\
  }\bibfield  {title} {\bibinfo {title} {Structure of avalanches and breakdown
  of simple scaling in the abelian sandpile model in one dimension},\ }\href
  {https://doi.org/10.1103/PhysRevE.52.4804} {\bibfield  {journal} {\bibinfo
  {journal} {Physical Review E}\ }\textbf {\bibinfo {volume} {52}},\ \bibinfo
  {pages} {4804} (\bibinfo {year} {1995})}\BibitemShut {NoStop}%
\bibitem [{\citenamefont {Levine}(2011)}]{levine2011sandpile}%
  \BibitemOpen
  \bibfield  {author} {\bibinfo {author} {\bibfnamefont {L.}~\bibnamefont
  {Levine}},\ }\bibfield  {title} {\bibinfo {title} {Sandpile groups and
  spanning trees of directed line graphs},\ }\href
  {https://doi.org/10.1016/j.jcta.2010.04.001} {\bibfield  {journal} {\bibinfo
  {journal} {Journal of Combinatorial Theory, Series A}\ }\textbf {\bibinfo
  {volume} {118}},\ \bibinfo {pages} {350} (\bibinfo {year}
  {2011})}\BibitemShut {NoStop}%
\bibitem [{\citenamefont {Dhar}(1999)}]{dhar1999abelian}%
  \BibitemOpen
  \bibfield  {author} {\bibinfo {author} {\bibfnamefont {D.}~\bibnamefont
  {Dhar}},\ }\bibfield  {title} {\bibinfo {title} {The abelian sandpile and
  related models},\ }\href {https://doi.org/10.1016/S0378-4371(98)00493-2}
  {\bibfield  {journal} {\bibinfo  {journal} {Physica A: Statistical Mechanics
  and its Applications}\ }\textbf {\bibinfo {volume} {263}},\ \bibinfo {pages}
  {4} (\bibinfo {year} {1999})}\BibitemShut {NoStop}%
\bibitem [{\citenamefont {Dhar}(2006)}]{dhar2006theoretical}%
  \BibitemOpen
  \bibfield  {author} {\bibinfo {author} {\bibfnamefont {D.}~\bibnamefont
  {Dhar}},\ }\bibfield  {title} {\bibinfo {title} {Theoretical studies of
  self-organized criticality},\ }\href
  {https://doi.org/10.1016/j.physa.2006.04.004} {\bibfield  {journal} {\bibinfo
   {journal} {Physica A: Statistical Mechanics and its Applications}\ }\textbf
  {\bibinfo {volume} {369}},\ \bibinfo {pages} {29} (\bibinfo {year}
  {2006})}\BibitemShut {NoStop}%
\bibitem [{\citenamefont {Manna}(2023)}]{manna2023nonstationary}%
  \BibitemOpen
  \bibfield  {author} {\bibinfo {author} {\bibfnamefont {S.}~\bibnamefont
  {Manna}},\ }\bibfield  {title} {\bibinfo {title} {Nonstationary but
  quasisteady states in self-organized criticality},\ }\href
  {https://doi.org/10.1103/PhysRevE.107.044113} {\bibfield  {journal} {\bibinfo
   {journal} {Physical Review E}\ }\textbf {\bibinfo {volume} {107}},\ \bibinfo
  {pages} {044113} (\bibinfo {year} {2023})}\BibitemShut {NoStop}%
\bibitem [{\citenamefont {Najafi}\ \emph
  {et~al.}(2020{\natexlab{a}})\citenamefont {Najafi}, \citenamefont
  {Moghaddam}, \citenamefont {Samadpour},\ and\ \citenamefont
  {Araújo}}]{Najafi_2020}%
  \BibitemOpen
  \bibfield  {author} {\bibinfo {author} {\bibfnamefont {M.~N.}\ \bibnamefont
  {Najafi}}, \bibinfo {author} {\bibfnamefont {Z.}~\bibnamefont {Moghaddam}},
  \bibinfo {author} {\bibfnamefont {M.}~\bibnamefont {Samadpour}},\ and\
  \bibinfo {author} {\bibfnamefont {N.~A.~M.}\ \bibnamefont {Araújo}},\
  }\bibfield  {title} {\bibinfo {title} {Invasion sandpile model},\ }\href
  {https://doi.org/10.1088/1742-5468/ab96b4} {\bibfield  {journal} {\bibinfo
  {journal} {Journal of Statistical Mechanics: Theory and Experiment}\ }\textbf
  {\bibinfo {volume} {2020}},\ \bibinfo {pages} {073205} (\bibinfo {year}
  {2020}{\natexlab{a}})}\BibitemShut {NoStop}%
\bibitem [{\citenamefont {Lise}\ and\ \citenamefont
  {Paczuski}(2001)}]{Lise2001earthquake}%
  \BibitemOpen
  \bibfield  {author} {\bibinfo {author} {\bibfnamefont {S.}~\bibnamefont
  {Lise}}\ and\ \bibinfo {author} {\bibfnamefont {M.}~\bibnamefont
  {Paczuski}},\ }\bibfield  {title} {\bibinfo {title} {Scaling in a
  nonconservative earthquake model of self-organized criticality},\ }\href
  {https://doi.org/10.1103/PhysRevE.64.046111} {\bibfield  {journal} {\bibinfo
  {journal} {Phys. Rev. E}\ }\textbf {\bibinfo {volume} {64}},\ \bibinfo
  {pages} {046111} (\bibinfo {year} {2001})}\BibitemShut {NoStop}%
\bibitem [{\citenamefont {Najafi}\ \emph {et~al.}(2016)\citenamefont {Najafi},
  \citenamefont {Ghaedi},\ and\ \citenamefont
  {Moghimi-Araghi}}]{NAJAFI2016102}%
  \BibitemOpen
  \bibfield  {author} {\bibinfo {author} {\bibfnamefont {M.}~\bibnamefont
  {Najafi}}, \bibinfo {author} {\bibfnamefont {M.}~\bibnamefont {Ghaedi}},\
  and\ \bibinfo {author} {\bibfnamefont {S.}~\bibnamefont {Moghimi-Araghi}},\
  }\bibfield  {title} {\bibinfo {title} {Water propagation in two-dimensional
  petroleum reservoirs},\ }\href
  {https://doi.org/https://doi.org/10.1016/j.physa.2015.10.100} {\bibfield
  {journal} {\bibinfo  {journal} {Physica A: Statistical Mechanics and its
  Applications}\ }\textbf {\bibinfo {volume} {445}},\ \bibinfo {pages} {102}
  (\bibinfo {year} {2016})}\BibitemShut {NoStop}%
\bibitem [{\citenamefont {Dashti-Naserabadi}\ and\ \citenamefont
  {Najafi}(2015)}]{dashti2015statistical}%
  \BibitemOpen
  \bibfield  {author} {\bibinfo {author} {\bibfnamefont {H.}~\bibnamefont
  {Dashti-Naserabadi}}\ and\ \bibinfo {author} {\bibfnamefont {M.}~\bibnamefont
  {Najafi}},\ }\bibfield  {title} {\bibinfo {title} {Statistical investigation
  of the cross sections of wave clusters in the three-dimensional
  bak-tang-wiesenfeld model},\ }\href
  {https://doi.org/10.1103/PhysRevE.91.052145} {\bibfield  {journal} {\bibinfo
  {journal} {Physical Review E}\ }\textbf {\bibinfo {volume} {91}},\ \bibinfo
  {pages} {052145} (\bibinfo {year} {2015})}\BibitemShut {NoStop}%
\bibitem [{\citenamefont {Saberi}(2015)}]{saberi2015recent}%
  \BibitemOpen
  \bibfield  {author} {\bibinfo {author} {\bibfnamefont {A.~A.}\ \bibnamefont
  {Saberi}},\ }\bibfield  {title} {\bibinfo {title} {Recent advances in
  percolation theory and its applications},\ }\href
  {https://doi.org/10.1016/j.physrep.2015.03.003} {\bibfield  {journal}
  {\bibinfo  {journal} {Physics Reports}\ }\textbf {\bibinfo {volume} {578}},\
  \bibinfo {pages} {1} (\bibinfo {year} {2015})}\BibitemShut {NoStop}%
\bibitem [{\citenamefont {De~Noronha}\ \emph {et~al.}(2018)\citenamefont
  {De~Noronha}, \citenamefont {Moreira}, \citenamefont {Vieira}, \citenamefont
  {Herrmann}, \citenamefont {Andrade~Jr},\ and\ \citenamefont
  {Carmona}}]{de2018percolation}%
  \BibitemOpen
  \bibfield  {author} {\bibinfo {author} {\bibfnamefont {A.~W.}\ \bibnamefont
  {De~Noronha}}, \bibinfo {author} {\bibfnamefont {A.~A.}\ \bibnamefont
  {Moreira}}, \bibinfo {author} {\bibfnamefont {A.~P.}\ \bibnamefont {Vieira}},
  \bibinfo {author} {\bibfnamefont {H.~J.}\ \bibnamefont {Herrmann}}, \bibinfo
  {author} {\bibfnamefont {J.~S.}\ \bibnamefont {Andrade~Jr}},\ and\ \bibinfo
  {author} {\bibfnamefont {H.~A.}\ \bibnamefont {Carmona}},\ }\bibfield
  {title} {\bibinfo {title} {Percolation on an isotropically directed
  lattice},\ }\href {https://doi.org/10.1103/PhysRevE.98.062116} {\bibfield
  {journal} {\bibinfo  {journal} {Physical Review E}\ }\textbf {\bibinfo
  {volume} {98}},\ \bibinfo {pages} {062116} (\bibinfo {year}
  {2018})}\BibitemShut {NoStop}%
\bibitem [{\citenamefont {Li}\ \emph {et~al.}(2016)\citenamefont {Li},
  \citenamefont {Wang}, \citenamefont {Liu},\ and\ \citenamefont
  {Hu}}]{li2016many}%
  \BibitemOpen
  \bibfield  {author} {\bibinfo {author} {\bibfnamefont {H.}~\bibnamefont
  {Li}}, \bibinfo {author} {\bibfnamefont {J.}~\bibnamefont {Wang}}, \bibinfo
  {author} {\bibfnamefont {X.-J.}\ \bibnamefont {Liu}},\ and\ \bibinfo {author}
  {\bibfnamefont {H.}~\bibnamefont {Hu}},\ }\bibfield  {title} {\bibinfo
  {title} {Many-body localization in ising models with random long-range
  interactions},\ }\href {https://doi.org/10.1103/PhysRevA.94.063625}
  {\bibfield  {journal} {\bibinfo  {journal} {Physical Review A}\ }\textbf
  {\bibinfo {volume} {94}},\ \bibinfo {pages} {063625} (\bibinfo {year}
  {2016})}\BibitemShut {NoStop}%
\bibitem [{\citenamefont {Barab{\'a}si}\ \emph {et~al.}(1999)\citenamefont
  {Barab{\'a}si}, \citenamefont {Albert},\ and\ \citenamefont
  {Jeong}}]{barabasi1999mean}%
  \BibitemOpen
  \bibfield  {author} {\bibinfo {author} {\bibfnamefont {A.-L.}\ \bibnamefont
  {Barab{\'a}si}}, \bibinfo {author} {\bibfnamefont {R.}~\bibnamefont
  {Albert}},\ and\ \bibinfo {author} {\bibfnamefont {H.}~\bibnamefont
  {Jeong}},\ }\bibfield  {title} {\bibinfo {title} {Mean-field theory for
  scale-free random networks},\ }\href
  {https://doi.org/10.1016/S0378-4371(99)00291-5} {\bibfield  {journal}
  {\bibinfo  {journal} {Physica A: Statistical Mechanics and its Applications}\
  }\textbf {\bibinfo {volume} {272}},\ \bibinfo {pages} {173} (\bibinfo {year}
  {1999})}\BibitemShut {NoStop}%
\bibitem [{\citenamefont {Christensen}\ and\ \citenamefont
  {Olami}(1993)}]{christensen1993sandpile}%
  \BibitemOpen
  \bibfield  {author} {\bibinfo {author} {\bibfnamefont {K.}~\bibnamefont
  {Christensen}}\ and\ \bibinfo {author} {\bibfnamefont {Z.}~\bibnamefont
  {Olami}},\ }\bibfield  {title} {\bibinfo {title} {Sandpile models with and
  without an underlying spatial structure},\ }\href
  {https://doi.org/10.1103/PhysRevE.48.3361} {\bibfield  {journal} {\bibinfo
  {journal} {Physical Review E}\ }\textbf {\bibinfo {volume} {48}},\ \bibinfo
  {pages} {3361} (\bibinfo {year} {1993})}\BibitemShut {NoStop}%
\bibitem [{\citenamefont {Moosavi}\ and\ \citenamefont
  {Montakhab}(2014)}]{moosavi2014mean}%
  \BibitemOpen
  \bibfield  {author} {\bibinfo {author} {\bibfnamefont {S.~A.}\ \bibnamefont
  {Moosavi}}\ and\ \bibinfo {author} {\bibfnamefont {A.}~\bibnamefont
  {Montakhab}},\ }\bibfield  {title} {\bibinfo {title} {Mean-field behavior as
  a result of noisy local dynamics in self-organized criticality: Neuroscience
  implications},\ }\href {https://doi.org/10.1103/PhysRevE.89.052139}
  {\bibfield  {journal} {\bibinfo  {journal} {Physical Review E}\ }\textbf
  {\bibinfo {volume} {89}},\ \bibinfo {pages} {052139} (\bibinfo {year}
  {2014})}\BibitemShut {NoStop}%
\bibitem [{\citenamefont {Rahimi-Majd}\ \emph {et~al.}(2021)\citenamefont
  {Rahimi-Majd}, \citenamefont {Seifi}, \citenamefont {de~Arcangelis},\ and\
  \citenamefont {Najafi}}]{rahimi2021role}%
  \BibitemOpen
  \bibfield  {author} {\bibinfo {author} {\bibfnamefont {M.}~\bibnamefont
  {Rahimi-Majd}}, \bibinfo {author} {\bibfnamefont {M.}~\bibnamefont {Seifi}},
  \bibinfo {author} {\bibfnamefont {L.}~\bibnamefont {de~Arcangelis}},\ and\
  \bibinfo {author} {\bibfnamefont {M.}~\bibnamefont {Najafi}},\ }\bibfield
  {title} {\bibinfo {title} {Role of anaxonic local neurons in the crossover to
  continuously varying exponents for avalanche activity},\ }\href
  {https://doi.org/0.1103/PhysRevE.103.042402} {\bibfield  {journal} {\bibinfo
  {journal} {Physical Review E}\ }\textbf {\bibinfo {volume} {103}},\ \bibinfo
  {pages} {042402} (\bibinfo {year} {2021})}\BibitemShut {NoStop}%
\bibitem [{\citenamefont {Strogatz}(2001)}]{strogatz2001exploring}%
  \BibitemOpen
  \bibfield  {author} {\bibinfo {author} {\bibfnamefont {S.~H.}\ \bibnamefont
  {Strogatz}},\ }\bibfield  {title} {\bibinfo {title} {Exploring complex
  networks},\ }\href {https://doi.org/10.1038/35065725} {\bibfield  {journal}
  {\bibinfo  {journal} {nature}\ }\textbf {\bibinfo {volume} {410}},\ \bibinfo
  {pages} {268} (\bibinfo {year} {2001})}\BibitemShut {NoStop}%
\bibitem [{\citenamefont {Landi}\ \emph {et~al.}(2018)\citenamefont {Landi},
  \citenamefont {Minoarivelo}, \citenamefont {Br{\"a}nnstr{\"o}m},
  \citenamefont {Hui},\ and\ \citenamefont {Dieckmann}}]{landi2018complexity}%
  \BibitemOpen
  \bibfield  {author} {\bibinfo {author} {\bibfnamefont {P.}~\bibnamefont
  {Landi}}, \bibinfo {author} {\bibfnamefont {H.~O.}\ \bibnamefont
  {Minoarivelo}}, \bibinfo {author} {\bibfnamefont {{\AA}.}~\bibnamefont
  {Br{\"a}nnstr{\"o}m}}, \bibinfo {author} {\bibfnamefont {C.}~\bibnamefont
  {Hui}},\ and\ \bibinfo {author} {\bibfnamefont {U.}~\bibnamefont
  {Dieckmann}},\ }\bibfield  {title} {\bibinfo {title} {Complexity and
  stability of ecological networks: a review of the theory},\ }\href
  {https://doi.org/10.1007/s10144-018-0628-3} {\bibfield  {journal} {\bibinfo
  {journal} {Population ecology}\ }\textbf {\bibinfo {volume} {60}},\ \bibinfo
  {pages} {319} (\bibinfo {year} {2018})}\BibitemShut {NoStop}%
\bibitem [{\citenamefont {Frasco}\ \emph {et~al.}(2014)\citenamefont {Frasco},
  \citenamefont {Sun}, \citenamefont {Rozenfeld},\ and\ \citenamefont
  {Ben-Avraham}}]{frasco2014spatially}%
  \BibitemOpen
  \bibfield  {author} {\bibinfo {author} {\bibfnamefont {G.~F.}\ \bibnamefont
  {Frasco}}, \bibinfo {author} {\bibfnamefont {J.}~\bibnamefont {Sun}},
  \bibinfo {author} {\bibfnamefont {H.~D.}\ \bibnamefont {Rozenfeld}},\ and\
  \bibinfo {author} {\bibfnamefont {D.}~\bibnamefont {Ben-Avraham}},\
  }\bibfield  {title} {\bibinfo {title} {Spatially distributed social complex
  networks},\ }\href {https://doi.org/10.1103/PhysRevX.4.011008} {\bibfield
  {journal} {\bibinfo  {journal} {Physical Review X}\ }\textbf {\bibinfo
  {volume} {4}},\ \bibinfo {pages} {011008} (\bibinfo {year}
  {2014})}\BibitemShut {NoStop}%
\bibitem [{\citenamefont {Sol{\'e}}\ and\ \citenamefont
  {Valverde}(2004)}]{sole2004information}%
  \BibitemOpen
  \bibfield  {author} {\bibinfo {author} {\bibfnamefont {R.~V.}\ \bibnamefont
  {Sol{\'e}}}\ and\ \bibinfo {author} {\bibfnamefont {S.}~\bibnamefont
  {Valverde}},\ }\bibfield  {title} {\bibinfo {title} {Information theory of
  complex networks: on evolution and architectural constraints},\ }in\ \href
  {https://doi.org/10.1007/978-3-540-44485-5_9} {\emph {\bibinfo {booktitle}
  {Complex networks}}}\ (\bibinfo  {publisher} {Springer},\ \bibinfo {year}
  {2004})\ pp.\ \bibinfo {pages} {189--207}\BibitemShut {NoStop}%
\bibitem [{\citenamefont {Newman}(2018)}]{newman2018networks}%
  \BibitemOpen
  \bibfield  {author} {\bibinfo {author} {\bibfnamefont {M.}~\bibnamefont
  {Newman}},\ }\href@noop {} {\emph {\bibinfo {title} {Networks}}}\ (\bibinfo
  {publisher} {Oxford university press},\ \bibinfo {year} {2018})\BibitemShut
  {NoStop}%
\bibitem [{\citenamefont {Pan}\ \emph {et~al.}(2007)\citenamefont {Pan},
  \citenamefont {Zhang}, \citenamefont {Yin},\ and\ \citenamefont
  {He}}]{pan2007sandpile}%
  \BibitemOpen
  \bibfield  {author} {\bibinfo {author} {\bibfnamefont {G.-J.}\ \bibnamefont
  {Pan}}, \bibinfo {author} {\bibfnamefont {D.-M.}\ \bibnamefont {Zhang}},
  \bibinfo {author} {\bibfnamefont {Y.-P.}\ \bibnamefont {Yin}},\ and\ \bibinfo
  {author} {\bibfnamefont {M.-H.}\ \bibnamefont {He}},\ }\bibfield  {title}
  {\bibinfo {title} {Sandpile on directed small-world networks},\ }\href
  {https://doi.org/10.1016/j.physa.2007.04.113} {\bibfield  {journal} {\bibinfo
   {journal} {Physica A: Statistical Mechanics and its Applications}\ }\textbf
  {\bibinfo {volume} {383}},\ \bibinfo {pages} {435} (\bibinfo {year}
  {2007})}\BibitemShut {NoStop}%
\bibitem [{\citenamefont {Bhaumik}\ and\ \citenamefont
  {Santra}(2018)}]{bhaumik2018stochastic}%
  \BibitemOpen
  \bibfield  {author} {\bibinfo {author} {\bibfnamefont {H.}~\bibnamefont
  {Bhaumik}}\ and\ \bibinfo {author} {\bibfnamefont {S.}~\bibnamefont
  {Santra}},\ }\bibfield  {title} {\bibinfo {title} {Stochastic sandpile model
  on small-world networks: Scaling and crossover},\ }\href
  {https://doi.org/10.1016/j.physa.2018.08.003} {\bibfield  {journal} {\bibinfo
   {journal} {Physica A: Statistical Mechanics and its Applications}\ }\textbf
  {\bibinfo {volume} {511}},\ \bibinfo {pages} {358} (\bibinfo {year}
  {2018})}\BibitemShut {NoStop}%
\bibitem [{\citenamefont {Lelarge}(2012)}]{lelarge2012diffusion}%
  \BibitemOpen
  \bibfield  {author} {\bibinfo {author} {\bibfnamefont {M.}~\bibnamefont
  {Lelarge}},\ }\bibfield  {title} {\bibinfo {title} {Diffusion and cascading
  behavior in random networks},\ }\href
  {https://doi.org/10.1016/j.geb.2012.03.009} {\bibfield  {journal} {\bibinfo
  {journal} {Games and Economic Behavior}\ }\textbf {\bibinfo {volume} {75}},\
  \bibinfo {pages} {752} (\bibinfo {year} {2012})}\BibitemShut {NoStop}%
\bibitem [{\citenamefont {Lee}\ \emph {et~al.}(2004)\citenamefont {Lee},
  \citenamefont {Goh}, \citenamefont {Kahng},\ and\ \citenamefont
  {Kim}}]{lee2004branching}%
  \BibitemOpen
  \bibfield  {author} {\bibinfo {author} {\bibfnamefont {D.~S.}\ \bibnamefont
  {Lee}}, \bibinfo {author} {\bibfnamefont {K.~I.}\ \bibnamefont {Goh}},
  \bibinfo {author} {\bibfnamefont {B.}~\bibnamefont {Kahng}},\ and\ \bibinfo
  {author} {\bibfnamefont {D.}~\bibnamefont {Kim}},\ }\bibfield  {title}
  {\bibinfo {title} {Branching process approach to avalanche dynamics on
  complex networks},\ }\href {https://doi.org/here} {\bibfield  {journal}
  {\bibinfo  {journal} {JOURNAL-KOREAN PHYSICAL SOCIETY}\ }\textbf {\bibinfo
  {volume} {44}},\ \bibinfo {pages} {633} (\bibinfo {year} {2004})}\BibitemShut
  {NoStop}%
\bibitem [{\citenamefont {Lee}\ \emph {et~al.}(2015)\citenamefont {Lee},
  \citenamefont {Kim}, \citenamefont {Lee},\ and\ \citenamefont
  {Kahng}}]{lee2015forest}%
  \BibitemOpen
  \bibfield  {author} {\bibinfo {author} {\bibfnamefont {D.}~\bibnamefont
  {Lee}}, \bibinfo {author} {\bibfnamefont {J.-Y.}\ \bibnamefont {Kim}},
  \bibinfo {author} {\bibfnamefont {J.}~\bibnamefont {Lee}},\ and\ \bibinfo
  {author} {\bibfnamefont {B.}~\bibnamefont {Kahng}},\ }\bibfield  {title}
  {\bibinfo {title} {Forest-fire model as a supercritical dynamic model in
  financial systems},\ }\href {https://doi.org/10.1103/PhysRevE.91.022806}
  {\bibfield  {journal} {\bibinfo  {journal} {Physical Review E}\ }\textbf
  {\bibinfo {volume} {91}},\ \bibinfo {pages} {022806} (\bibinfo {year}
  {2015})}\BibitemShut {NoStop}%
\bibitem [{\citenamefont {Goh}\ \emph {et~al.}(2003)\citenamefont {Goh},
  \citenamefont {Lee}, \citenamefont {Kahng},\ and\ \citenamefont
  {Kim}}]{goh2003sandpile}%
  \BibitemOpen
  \bibfield  {author} {\bibinfo {author} {\bibfnamefont {K.-I.}\ \bibnamefont
  {Goh}}, \bibinfo {author} {\bibfnamefont {D.-S.}\ \bibnamefont {Lee}},
  \bibinfo {author} {\bibfnamefont {B.}~\bibnamefont {Kahng}},\ and\ \bibinfo
  {author} {\bibfnamefont {D.}~\bibnamefont {Kim}},\ }\bibfield  {title}
  {\bibinfo {title} {Sandpile on scale-free networks},\ }\href
  {https://doi.org/10.1103/PhysRevLett.91.148701} {\bibfield  {journal}
  {\bibinfo  {journal} {Physical review letters}\ }\textbf {\bibinfo {volume}
  {91}},\ \bibinfo {pages} {148701} (\bibinfo {year} {2003})}\BibitemShut
  {NoStop}%
\bibitem [{\citenamefont {Najafi}(2014)}]{najafi2014bak}%
  \BibitemOpen
  \bibfield  {author} {\bibinfo {author} {\bibfnamefont {M.}~\bibnamefont
  {Najafi}},\ }\bibfield  {title} {\bibinfo {title} {Bak--tang--wiesenfeld
  model in the finite range random link lattice},\ }\href
  {https://doi.org/10.1016/j.physleta.2014.05.051} {\bibfield  {journal}
  {\bibinfo  {journal} {Physics Letters A}\ }\textbf {\bibinfo {volume}
  {378}},\ \bibinfo {pages} {2008} (\bibinfo {year} {2014})}\BibitemShut
  {NoStop}%
\bibitem [{\citenamefont {Najafi}\ and\ \citenamefont
  {Dashti-Naserabadi}(2018{\natexlab{a}})}]{najafi2018statistical}%
  \BibitemOpen
  \bibfield  {author} {\bibinfo {author} {\bibfnamefont {M.}~\bibnamefont
  {Najafi}}\ and\ \bibinfo {author} {\bibfnamefont {H.}~\bibnamefont
  {Dashti-Naserabadi}},\ }\bibfield  {title} {\bibinfo {title} {Statistical
  investigation of avalanches of three-dimensional small-world networks and
  their boundary and bulk cross-sections},\ }\href
  {https://doi.org/10.1103/PhysRevE.97.032108} {\bibfield  {journal} {\bibinfo
  {journal} {Physical Review E}\ }\textbf {\bibinfo {volume} {97}},\ \bibinfo
  {pages} {032108} (\bibinfo {year} {2018}{\natexlab{a}})}\BibitemShut
  {NoStop}%
\bibitem [{\citenamefont {Najafi}\ and\ \citenamefont
  {Dashti-Naserabadi}(2018{\natexlab{b}})}]{najafi2018sandpile}%
  \BibitemOpen
  \bibfield  {author} {\bibinfo {author} {\bibfnamefont {M.}~\bibnamefont
  {Najafi}}\ and\ \bibinfo {author} {\bibfnamefont {H.}~\bibnamefont
  {Dashti-Naserabadi}},\ }\bibfield  {title} {\bibinfo {title} {Sandpile on
  uncorrelated site-diluted percolation lattice; from three to two
  dimensions},\ }\href {https://doi.org/10.1088/1742-5468} {\bibfield
  {journal} {\bibinfo  {journal} {Journal of Statistical Mechanics: Theory and
  Experiment}\ }\textbf {\bibinfo {volume} {2018}},\ \bibinfo {pages} {023211}
  (\bibinfo {year} {2018}{\natexlab{b}})}\BibitemShut {NoStop}%
\bibitem [{\citenamefont {Najafi}\ \emph
  {et~al.}(2020{\natexlab{b}})\citenamefont {Najafi}, \citenamefont
  {Rahimi-Majd},\ and\ \citenamefont {Shirzad}}]{najafi2020avalanches}%
  \BibitemOpen
  \bibfield  {author} {\bibinfo {author} {\bibfnamefont {M.}~\bibnamefont
  {Najafi}}, \bibinfo {author} {\bibfnamefont {M.}~\bibnamefont
  {Rahimi-Majd}},\ and\ \bibinfo {author} {\bibfnamefont {T.}~\bibnamefont
  {Shirzad}},\ }\bibfield  {title} {\bibinfo {title} {Avalanches on the complex
  network of rigan earthquake},\ }\href
  {https://doi.org/10.1209/0295-5075/130/20001} {\bibfield  {journal} {\bibinfo
   {journal} {Europhysics Letters}\ }\textbf {\bibinfo {volume} {130}},\
  \bibinfo {pages} {20001} (\bibinfo {year} {2020}{\natexlab{b}})}\BibitemShut
  {NoStop}%
\bibitem [{\citenamefont {Najafi}\ \emph
  {et~al.}(2020{\natexlab{c}})\citenamefont {Najafi}, \citenamefont
  {Cheraghalizadeh}, \citenamefont {Lukovi{\'c}},\ and\ \citenamefont
  {Herrmann}}]{najafi2020geometry}%
  \BibitemOpen
  \bibfield  {author} {\bibinfo {author} {\bibfnamefont {M.~N.}\ \bibnamefont
  {Najafi}}, \bibinfo {author} {\bibfnamefont {J.}~\bibnamefont
  {Cheraghalizadeh}}, \bibinfo {author} {\bibfnamefont {M.}~\bibnamefont
  {Lukovi{\'c}}},\ and\ \bibinfo {author} {\bibfnamefont {H.~J.}\ \bibnamefont
  {Herrmann}},\ }\bibfield  {title} {\bibinfo {title} {Geometry-induced
  nonequilibrium phase transition in sandpiles},\ }\href
  {https://doi.org/10.1103/PhysRevE.101.032116} {\bibfield  {journal} {\bibinfo
   {journal} {Physical Review E}\ }\textbf {\bibinfo {volume} {101}},\ \bibinfo
  {pages} {032116} (\bibinfo {year} {2020}{\natexlab{c}})}\BibitemShut
  {NoStop}%
\bibitem [{\citenamefont {Cheraghalizadeh}\ \emph {et~al.}(2017)\citenamefont
  {Cheraghalizadeh}, \citenamefont {Najafi}, \citenamefont
  {Dashti-Naserabadi},\ and\ \citenamefont
  {Mohammadzadeh}}]{cheraghalizadeh2017mapping}%
  \BibitemOpen
  \bibfield  {author} {\bibinfo {author} {\bibfnamefont {J.}~\bibnamefont
  {Cheraghalizadeh}}, \bibinfo {author} {\bibfnamefont {M.}~\bibnamefont
  {Najafi}}, \bibinfo {author} {\bibfnamefont {H.}~\bibnamefont
  {Dashti-Naserabadi}},\ and\ \bibinfo {author} {\bibfnamefont
  {H.}~\bibnamefont {Mohammadzadeh}},\ }\bibfield  {title} {\bibinfo {title}
  {Mapping of the bak, tang, and wiesenfeld sandpile model on a two-dimensional
  ising-correlated percolation lattice to the two-dimensional self-avoiding
  random walk},\ }\href {https://doi.org/10.1103/PhysRevE.96.052127} {\bibfield
   {journal} {\bibinfo  {journal} {Physical Review E}\ }\textbf {\bibinfo
  {volume} {96}},\ \bibinfo {pages} {052127} (\bibinfo {year}
  {2017})}\BibitemShut {NoStop}%
\bibitem [{\citenamefont {Najafi}(2016)}]{najafi2016bak}%
  \BibitemOpen
  \bibfield  {author} {\bibinfo {author} {\bibfnamefont {M.}~\bibnamefont
  {Najafi}},\ }\bibfield  {title} {\bibinfo {title} {Bak--tang--wiesenfeld
  model on the square site-percolation lattice},\ }\href
  {https://doi.org/10.1088/1751-8113/49/33/335003} {\bibfield  {journal}
  {\bibinfo  {journal} {Journal of Physics A: Mathematical and Theoretical}\
  }\textbf {\bibinfo {volume} {49}},\ \bibinfo {pages} {335003} (\bibinfo
  {year} {2016})}\BibitemShut {NoStop}%
\bibitem [{\citenamefont {Corral}(2004)}]{corral2004long}%
  \BibitemOpen
  \bibfield  {author} {\bibinfo {author} {\bibfnamefont {{\'A}.}~\bibnamefont
  {Corral}},\ }\bibfield  {title} {\bibinfo {title} {Long-term clustering,
  scaling, and universality in the temporal occurrence of earthquakes},\ }\href
  {https://doi.org/10.1103/PhysRevLett.92.108501} {\bibfield  {journal}
  {\bibinfo  {journal} {Physical Review Letters}\ }\textbf {\bibinfo {volume}
  {92}},\ \bibinfo {pages} {108501} (\bibinfo {year} {2004})}\BibitemShut
  {NoStop}%
\bibitem [{\citenamefont {Najafi}\ \emph
  {et~al.}(2020{\natexlab{d}})\citenamefont {Najafi}, \citenamefont
  {Rahimi-Majd},\ and\ \citenamefont {Shirzad}}]{najafi2020avalanches2}%
  \BibitemOpen
  \bibfield  {author} {\bibinfo {author} {\bibfnamefont {M.}~\bibnamefont
  {Najafi}}, \bibinfo {author} {\bibfnamefont {M.}~\bibnamefont
  {Rahimi-Majd}},\ and\ \bibinfo {author} {\bibfnamefont {T.}~\bibnamefont
  {Shirzad}},\ }\bibfield  {title} {\bibinfo {title} {Avalanches on the complex
  network of rigan earthquake},\ }\href@noop {} {\bibfield  {journal} {\bibinfo
   {journal} {Europhysics Letters}\ }\textbf {\bibinfo {volume} {130}},\
  \bibinfo {pages} {20001} (\bibinfo {year} {2020}{\natexlab{d}})}\BibitemShut
  {NoStop}%
\bibitem [{\citenamefont {Hesse}\ and\ \citenamefont
  {Gross}(2014)}]{hesse2014self}%
  \BibitemOpen
  \bibfield  {author} {\bibinfo {author} {\bibfnamefont {J.}~\bibnamefont
  {Hesse}}\ and\ \bibinfo {author} {\bibfnamefont {T.}~\bibnamefont {Gross}},\
  }\bibfield  {title} {\bibinfo {title} {Self-organized criticality as a
  fundamental property of neural systems},\ }\href
  {https://doi.org/10.3389/fnsys.2014.00166} {\bibfield  {journal} {\bibinfo
  {journal} {Frontiers in systems neuroscience}\ }\textbf {\bibinfo {volume}
  {8}},\ \bibinfo {pages} {166} (\bibinfo {year} {2014})}\BibitemShut {NoStop}%
\bibitem [{\citenamefont {Fazli}\ and\ \citenamefont
  {Azimi-Tafreshi}(2022)}]{fazli2022emergence}%
  \BibitemOpen
  \bibfield  {author} {\bibinfo {author} {\bibfnamefont {D.}~\bibnamefont
  {Fazli}}\ and\ \bibinfo {author} {\bibfnamefont {N.}~\bibnamefont
  {Azimi-Tafreshi}},\ }\bibfield  {title} {\bibinfo {title} {Emergence of
  oscillations in fixed-energy sandpile models on complex networks},\ }\href
  {https://doi.org/10.1103/PhysRevE.105.014303} {\bibfield  {journal} {\bibinfo
   {journal} {Physical Review E}\ }\textbf {\bibinfo {volume} {105}},\ \bibinfo
  {pages} {014303} (\bibinfo {year} {2022})}\BibitemShut {NoStop}%
\bibitem [{\citenamefont {Cajueiro}\ and\ \citenamefont
  {Andrade}(2010)}]{cajueiro2010controlling}%
  \BibitemOpen
  \bibfield  {author} {\bibinfo {author} {\bibfnamefont {D.~O.}\ \bibnamefont
  {Cajueiro}}\ and\ \bibinfo {author} {\bibfnamefont {R.~F.}\ \bibnamefont
  {Andrade}},\ }\bibfield  {title} {\bibinfo {title} {Controlling
  self-organized criticality in complex networks},\ }\href
  {https://doi.org/here} {\bibfield  {journal} {\bibinfo  {journal} {The
  European Physical Journal B}\ }\textbf {\bibinfo {volume} {77}},\ \bibinfo
  {pages} {291} (\bibinfo {year} {2010})}\BibitemShut {NoStop}%
\bibitem [{\citenamefont {Liu}\ \emph {et~al.}(2012)\citenamefont {Liu},
  \citenamefont {Wang}, \citenamefont {Lai},\ and\ \citenamefont
  {Wang}}]{liu2012cascading}%
  \BibitemOpen
  \bibfield  {author} {\bibinfo {author} {\bibfnamefont {R.~R.}\ \bibnamefont
  {Liu}}, \bibinfo {author} {\bibfnamefont {W.~X.}\ \bibnamefont {Wang}},
  \bibinfo {author} {\bibfnamefont {Y.-C.}\ \bibnamefont {Lai}},\ and\ \bibinfo
  {author} {\bibfnamefont {B.~H.}\ \bibnamefont {Wang}},\ }\bibfield  {title}
  {\bibinfo {title} {Cascading dynamics on random networks: Crossover in phase
  transition},\ }\href {https://doi.org/10.1103/PhysRevE.85.026110} {\bibfield
  {journal} {\bibinfo  {journal} {Physical Review E-Statistical, Nonlinear, and
  Soft Matter Physics}\ }\textbf {\bibinfo {volume} {85}},\ \bibinfo {pages}
  {026110} (\bibinfo {year} {2012})}\BibitemShut {NoStop}%
\bibitem [{\citenamefont {Bhaumik}(2018)}]{bhaumik2018conserved}%
  \BibitemOpen
  \bibfield  {author} {\bibinfo {author} {\bibfnamefont {H.}~\bibnamefont
  {Bhaumik}},\ }\bibfield  {title} {\bibinfo {title} {Conserved manna model on
  barabasi--albert scale-free network},\ }\href
  {https://doi.org/10.1140/epjb/e2017-80602-9} {\bibfield  {journal} {\bibinfo
  {journal} {The European Physical Journal B}\ }\textbf {\bibinfo {volume}
  {91}},\ \bibinfo {pages} {1} (\bibinfo {year} {2018})}\BibitemShut {NoStop}%
\bibitem [{\citenamefont {Zonoozi}\ and\ \citenamefont
  {Dassanayake}(1997)}]{Zonoozi_622908}%
  \BibitemOpen
  \bibfield  {author} {\bibinfo {author} {\bibfnamefont {M.}~\bibnamefont
  {Zonoozi}}\ and\ \bibinfo {author} {\bibfnamefont {P.}~\bibnamefont
  {Dassanayake}},\ }\bibfield  {title} {\bibinfo {title} {User mobility
  modeling and characterization of mobility patterns},\ }\href
  {https://doi.org/10.1109/49.622908} {\bibfield  {journal} {\bibinfo
  {journal} {IEEE Journal on Selected Areas in Communications}\ }\textbf
  {\bibinfo {volume} {15}},\ \bibinfo {pages} {1239} (\bibinfo {year}
  {1997})}\BibitemShut {NoStop}%
\bibitem [{\citenamefont {Ruths}\ and\ \citenamefont
  {Ruths}(2016)}]{Ruths2016}%
  \BibitemOpen
  \bibfield  {author} {\bibinfo {author} {\bibfnamefont {D.}~\bibnamefont
  {Ruths}}\ and\ \bibinfo {author} {\bibfnamefont {J.}~\bibnamefont {Ruths}},\
  }\bibfield  {title} {\bibinfo {title} {Estimating the minimum control count
  of random network models},\ }\href {https://doi.org/10.1038/srep19818}
  {\bibfield  {journal} {\bibinfo  {journal} {Scientific Reports}\ }\textbf
  {\bibinfo {volume} {6}},\ \bibinfo {pages} {19818} (\bibinfo {year}
  {2016})}\BibitemShut {NoStop}%
\bibitem [{\citenamefont {Forsman}\ \emph {et~al.}(2014)\citenamefont
  {Forsman}, \citenamefont {Moll},\ and\ \citenamefont
  {Linder}}]{Forsman_PhysRevSTPER.10.020122}%
  \BibitemOpen
  \bibfield  {author} {\bibinfo {author} {\bibfnamefont {J.}~\bibnamefont
  {Forsman}}, \bibinfo {author} {\bibfnamefont {R.}~\bibnamefont {Moll}},\ and\
  \bibinfo {author} {\bibfnamefont {C.}~\bibnamefont {Linder}},\ }\bibfield
  {title} {\bibinfo {title} {Extending the theoretical framing for physics
  education research: An illustrative application of complexity science},\
  }\href {https://doi.org/10.1103/PhysRevSTPER.10.020122} {\bibfield  {journal}
  {\bibinfo  {journal} {Phys. Rev. ST Phys. Educ. Res.}\ }\textbf {\bibinfo
  {volume} {10}},\ \bibinfo {pages} {020122} (\bibinfo {year}
  {2014})}\BibitemShut {NoStop}%
\bibitem [{\citenamefont {Coon}\ \emph {et~al.}(2018)\citenamefont {Coon},
  \citenamefont {Dettmann},\ and\ \citenamefont {Georgiou}}]{coon2018entropy}%
  \BibitemOpen
  \bibfield  {author} {\bibinfo {author} {\bibfnamefont {J.~P.}\ \bibnamefont
  {Coon}}, \bibinfo {author} {\bibfnamefont {C.~P.}\ \bibnamefont {Dettmann}},\
  and\ \bibinfo {author} {\bibfnamefont {O.}~\bibnamefont {Georgiou}},\
  }\bibfield  {title} {\bibinfo {title} {Entropy of spatial network
  ensembles},\ }\href {https://doi.org/10.1103/PhysRevE.97.042319} {\bibfield
  {journal} {\bibinfo  {journal} {Physical Review E}\ }\textbf {\bibinfo
  {volume} {97}},\ \bibinfo {pages} {042319} (\bibinfo {year}
  {2018})}\BibitemShut {NoStop}%
\bibitem [{\citenamefont {Nesterov}(2024)}]{nesterov2024clustering}%
  \BibitemOpen
  \bibfield  {author} {\bibinfo {author} {\bibfnamefont {A.~I.}\ \bibnamefont
  {Nesterov}},\ }\bibfield  {title} {\bibinfo {title} {On clustering
  coefficients in complex networks},\ }\href@noop {} {\bibfield  {journal}
  {\bibinfo  {journal} {arXiv preprint arXiv:2401.02999}\ } (\bibinfo {year}
  {2024})}\BibitemShut {NoStop}%
\bibitem [{\citenamefont {Nattagh~Najafi}\ \emph {et~al.}(2023)\citenamefont
  {Nattagh~Najafi}, \citenamefont {Zayed},\ and\ \citenamefont
  {Nabavizadeh}}]{nattagh2023swarming}%
  \BibitemOpen
  \bibfield  {author} {\bibinfo {author} {\bibfnamefont {M.}~\bibnamefont
  {Nattagh~Najafi}}, \bibinfo {author} {\bibfnamefont {R.~M.~A.}\ \bibnamefont
  {Zayed}},\ and\ \bibinfo {author} {\bibfnamefont {S.~A.}\ \bibnamefont
  {Nabavizadeh}},\ }\bibfield  {title} {\bibinfo {title} {Swarming transition
  in super-diffusive self-propelled particles},\ }\href@noop {} {\bibfield
  {journal} {\bibinfo  {journal} {Entropy}\ }\textbf {\bibinfo {volume} {25}},\
  \bibinfo {pages} {817} (\bibinfo {year} {2023})}\BibitemShut {NoStop}%
\bibitem [{\citenamefont {Phillips}(1996)}]{phillips1996stretched}%
  \BibitemOpen
  \bibfield  {author} {\bibinfo {author} {\bibfnamefont {J.}~\bibnamefont
  {Phillips}},\ }\bibfield  {title} {\bibinfo {title} {Stretched exponential
  relaxation in molecular and electronic glasses},\ }\href@noop {} {\bibfield
  {journal} {\bibinfo  {journal} {Reports on Progress in Physics}\ }\textbf
  {\bibinfo {volume} {59}},\ \bibinfo {pages} {1133} (\bibinfo {year}
  {1996})}\BibitemShut {NoStop}%
\bibitem [{\citenamefont {Gomes}\ \emph {et~al.}(2008)\citenamefont {Gomes},
  \citenamefont {Combes},\ and\ \citenamefont
  {Dussauchoy}}]{gomes2008parameter}%
  \BibitemOpen
  \bibfield  {author} {\bibinfo {author} {\bibfnamefont {O.}~\bibnamefont
  {Gomes}}, \bibinfo {author} {\bibfnamefont {C.}~\bibnamefont {Combes}},\ and\
  \bibinfo {author} {\bibfnamefont {A.}~\bibnamefont {Dussauchoy}},\ }\bibfield
   {title} {\bibinfo {title} {Parameter estimation of the generalized gamma
  distribution},\ }\href {https://doi.org/10.1016/j.matcom.2008.02.006}
  {\bibfield  {journal} {\bibinfo  {journal} {Mathematics and Computers in
  Simulation}\ }\textbf {\bibinfo {volume} {79}},\ \bibinfo {pages} {955}
  (\bibinfo {year} {2008})}\BibitemShut {NoStop}%
\bibitem [{\citenamefont {Nolan}(2012)}]{nolan2012stable}%
  \BibitemOpen
  \bibfield  {author} {\bibinfo {author} {\bibfnamefont {J.~P.}\ \bibnamefont
  {Nolan}},\ }\href@noop {} {\emph {\bibinfo {title} {Stable distributions}}}\
  (\bibinfo {year} {2012})\BibitemShut {NoStop}%
\bibitem [{\citenamefont {Stacy}(1962)}]{stacy1962generalization}%
  \BibitemOpen
  \bibfield  {author} {\bibinfo {author} {\bibfnamefont {E.~W.}\ \bibnamefont
  {Stacy}},\ }\bibfield  {title} {\bibinfo {title} {A generalization of the
  gamma distribution},\ }\href@noop {} {\bibfield  {journal} {\bibinfo
  {journal} {The Annals of mathematical statistics}\ ,\ \bibinfo {pages}
  {1187}} (\bibinfo {year} {1962})}\BibitemShut {NoStop}%
\bibitem [{\citenamefont {Agarwal}\ and\ \citenamefont
  {Kalla}(1996)}]{agarwal1996generalized}%
  \BibitemOpen
  \bibfield  {author} {\bibinfo {author} {\bibfnamefont {S.~K.}\ \bibnamefont
  {Agarwal}}\ and\ \bibinfo {author} {\bibfnamefont {S.~L.}\ \bibnamefont
  {Kalla}},\ }\bibfield  {title} {\bibinfo {title} {A generalized gamma
  distribution and its application in reliability},\ }\href
  {https://doi.org/10.1080/03610929608831688} {\bibfield  {journal} {\bibinfo
  {journal} {Communications in Statistics - Theory and Methods}\ }\textbf
  {\bibinfo {volume} {25}},\ \bibinfo {pages} {201} (\bibinfo {year}
  {1996})}\BibitemShut {NoStop}%
\end{thebibliography}%

\newpage
\appendix
\renewcommand{\thesection}{\Alph{section}}
\renewcommand{\thefigure}{A.\arabic{figure}}
\setcounter{figure}{0}
\renewcommand{\theequation}{A.\arabic{equation}}
\setcounter{equation}{0}
\setcounter{page}{1}
\appendix

\section{Relation between GGD, and other distribution functions}\label{SEC:appendices}
In this section we explore some properties of generalized gamma distribution (GGD, represented by $G_{\alpha,\beta,\gamma}$), especially its connection to the other distributions. For $\alpha=0$, GGD corresponds to a \textit{stretched exponential} function which is related to the generalized Levy distribution function observed for many systems like the active version of sandpiles~\cite{nattagh2023swarming},and glass systems~\cite{phillips1996stretched}. In the other hand, for $\gamma=0$ this distribution function is power-law with the exponent $\alpha$. For these reasons, one may name GGD as an \textit{stretched exponential power-law} distribution. This distribution corresponds to a Le\'vy distribution function with the following parameters:
\begin{eqnarray}  
P_{\text{Le\'vy}}(x,\mu)&&=G_{-\frac{3}{2},\frac{c}{2},-1}(x-\mu)\\
&&=\sqrt{\frac{c}{2\pi}}\frac{1}{(x-\mu)^{\frac{3}{2}}}\exp\left[-\frac{c}{2(x-\mu)}\right],\nonumber
\end{eqnarray}
where $\mu$ is some average value. The GGD and the Lévy distribution are both flexible and powerful distributions used for modeling stochastic phenomena with specific characteristics such as heavy tails. Heavy tails imply that these distributions provide a higher probability for rare and extreme events, such as sudden jumps in financial prices or drastic changes in physical data. The GGD, due to its flexible structure, can mimic various distribution forms, including exponential, gamma, Weibull, and even Lévy distributions, highlighting its wide range of applications \cite{gomes2008parameter}. Consequently, the Lévy distribution can be considered as a special case of the GGD. The Lévy distribution is typically used to model sudden jumps and random dispersion in financial and physical processes~\cite{nolan2012stable}, while the GGD, with its more adaptable structure, can also model positive data, reliability modeling and lifetime distributions~\cite{stacy1962generalization}. Although the applications of these two distributions differ, they both play significant roles in analyzing data with heavy tails and substantial fluctuations\cite{agarwal1996generalized}.\\

GGD is also related to generalised gamma subordinator (GGS), with the distribution
\begin{equation}
P_{GGS}(x)=G_{-\sigma-1,\beta,1}(x)\propto \frac{e^{-\beta x}}{x^{1+\sigma}}.
\end{equation}

\section{Shannon Entropy}\label{App:Shannon}
\begin{figure*}[t]
  \centering
    {\includegraphics[width=\linewidth]{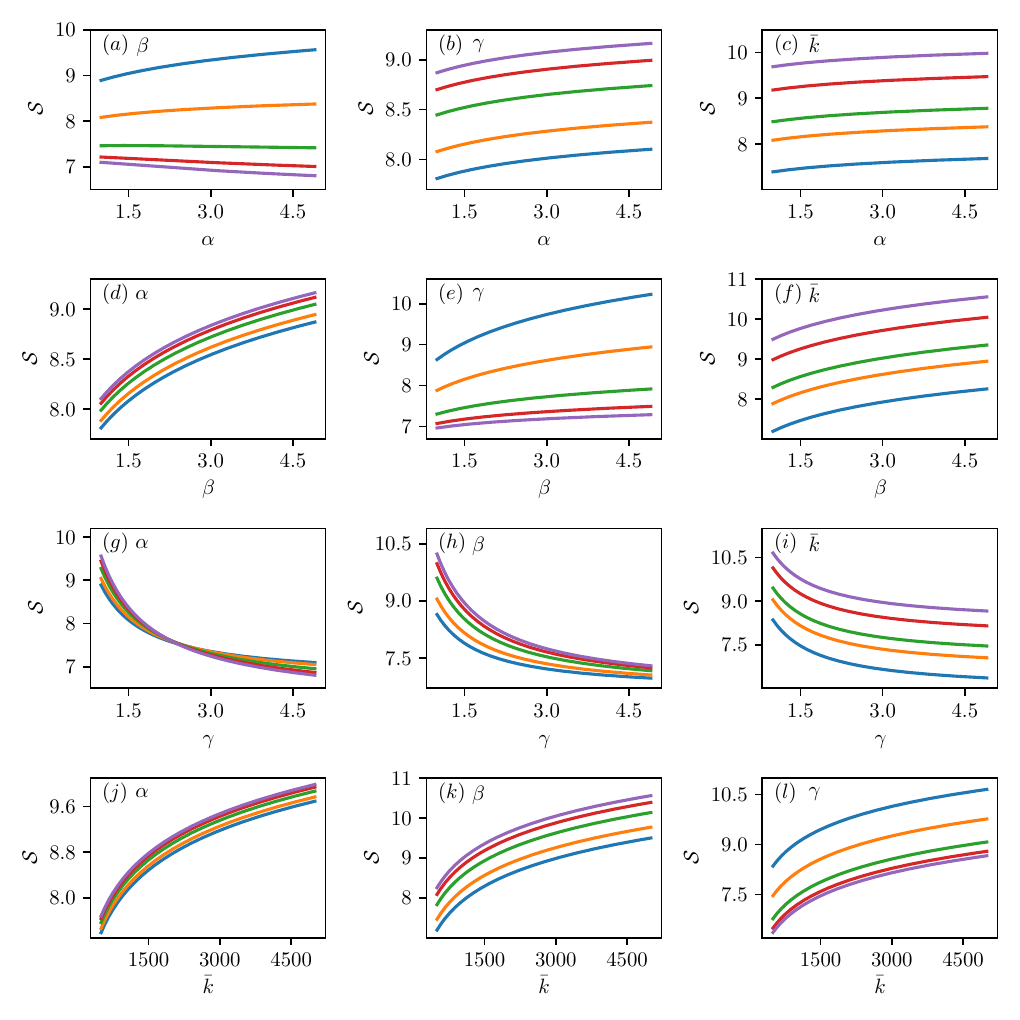}}    
    \caption{Analytical behavior of Shannon entropy associated with the degree centrality: The variation of entropy ($\mathcal{S}$) as a function of the parameters $\alpha$, $\beta$, $\gamma$, and the average degree of nodes ($\bar{k}$). In each subgraph, the varying parameter is indicated at the top, while the other two parameters are fixed ($\alpha$, $\beta$ and $\gamma$ are fixed to 1.5, while $\bar{k}$ is fixed to $1000$ if applicable). The exponents vary according to $(\alpha,\beta,\gamma)=1, 1.5,2.6, 3.8$, and $4.9$, while $\bar{k}=500,1000,1500, 3000$ and $5000$ (all from blue to purple) if applicable.}
    \phantomsection
    \label{fig:Entropy_Analytical}
\end{figure*}

In this section we review some properties of the Shannon entropy. The analytic form of the Shannon entropy associated with the nodes' degree with a distribution consistent with GGD is given in Eqs.~\ref{Eq:shannon2} and~\ref{Eq:shannon3}. Figure~\ref{fig:Entropy_Analytical} presents the Shannon entropy as a function of $\alpha$, $\beta$, $\gamma$ and $\bar{k}$, the parameters present in GGD. Figures~\ref{fig:Entropy_Analytical}a, ~\ref{fig:Entropy_Analytical}b and~\ref{fig:Entropy_Analytical}c give $\mathcal{S}$ in terms of $\alpha$, Figs.~\ref{fig:Entropy_Analytical}d, ~\ref{fig:Entropy_Analytical}e and~\ref{fig:Entropy_Analytical}f give $\mathcal{S}$ in terms of $\beta$, Figs.~\ref{fig:Entropy_Analytical}g, ~\ref{fig:Entropy_Analytical}h and~\ref{fig:Entropy_Analytical}i give $\mathcal{S}$ in terms of $\gamma$, and Figs.~\ref{fig:Entropy_Analytical}j, ~\ref{fig:Entropy_Analytical}k and~\ref{fig:Entropy_Analytical}l give $\mathcal{S}$ in terms of $\bar{k}$.\\

When entropy is plotted as a function of one of the three parameters $\alpha$, $\beta$, or the average degree of nodes ($\bar{k}$), while keeping the other two parameters fixed ($\alpha$, $\beta$, and $\gamma$ are set to 1.5, and $\bar{k}$ is fixed at 1000 if applicable), the entropy decreases for all values of $\gamma$ (Figs.~\ref{fig:Entropy_Analytical}g, \ref{fig:Entropy_Analytical}h, \ref{fig:Entropy_Analytical}i). This indicates that $\gamma$ plays a dominant role in reducing entropy, regardless of variations in the other parameters. In contrast, in other cases, entropy increases with respect to all parameters (Fig.~\ref{fig:Entropy_Analytical}) or, at most, exhibits a weak decreasing trend (as seen in Fig.~\ref{fig:Entropy_Analytical}a). This highlights that while $\alpha$, $\beta$, and $\bar{k}$ contribute to greater network diversity and complexity, $\gamma$ consistently reduces entropy, emphasizing its role in promoting homogeneity within the network structure. Overall, $\mathcal{S}$ generally increases with $\alpha$ (except for large $\gamma$ values, Fig.~\ref{fig:Entropy_Analytical}a), $\beta$, and $\bar{k}$, while it decreases with $\gamma$.

\end{document}